\newcommand{\diracslash}[1]{#1\llap{/\kern2pt}}

\newcommand{\be}{\begin{equation}}
	\newcommand{\ee}{\end{equation}}
\newcommand{\bea}{\begin{eqnarray}}
	\newcommand{\eea}{\end{eqnarray}}
\newcommand{\ba}[1]{\begin{array}{#1}}
	\newcommand{\ea}{\end{array}}

\newcommand{\bt}{\begin{tabular}}
	\newcommand{\et}{\end{tabular}}
\newcommand{\Tr}{{\rm Tr}}

\newcommand{\beas}{\begin{eqnarray*}}
	\newcommand{\eeas}{\end{eqnarray*}}

\RequirePackage[2020-02-02]{latexrelease}
\documentclass[preprint,prd,aps,amssymb,amsmath
,floats,nofootinbib,floatfix]{revtex4}

\DeclareSymbolFont{rsfs}{U}{rsfs}{m}{n}
\DeclareSymbolFontAlphabet{\mathrsfs}{rsfs}

\usepackage{graphicx}
\usepackage{multirow}
\usepackage{graphicx,epstopdf}

\usepackage{graphicx}
\usepackage{float}
\usepackage{enumitem}

\usepackage[utf8]{inputenc}
\usepackage[english]{babel}
\usepackage{hyperref}
\hypersetup{
	colorlinks=true,
	linkcolor=blue,
	filecolor=magenta,     
	citecolor=blue, 
	urlcolor=blue,
}
\usepackage{cleveref}

\renewcommand{\BibitemShut}[1]{}
\newcommand{\NITJ}{Department of Physics, Dr.\@ B.\@ R.\@ Ambedkar National Institute of Technology Jalandhar, Punjab 144011 India.}
\newcommand{\KENT}{Department of Physics, Kent State University, Kent, OH 44243 USA.}
\newcommand{\NIT}{Department of Physics, Dr.\@ B.\@ R.\@ Ambedkar National Institute of Technology Jalandhar, Punjab 144011 India.}

\begin{document}

\title{$\eta$ meson in strange magnetized matter} 
\author{Shivanshi Tiwari}
\email{shivanshitiwari4@gmail.com}
\affiliation{\NITJ}

 \author{Rajesh Kumar}
 \email{rajesh.sism@gmail.com}
 \affiliation{\KENT}

\author{Manisha Kumari}
\email{maniyadav93@gmail.com}
\affiliation{\NIT}

\author{Arvind Kumar}
\email{kumara@nitj.ac.in}
\affiliation{\NIT}

\def\be{\begin{equation}}
\def\ee{\end{equation}}
\def\bearr{\begin{eqnarray}}
\def\eearr{\end{eqnarray}}
\def\zbf#1{{\bf {#1}}}
\def\bfm#1{\mbox{\boldmath $#1$}}
\def\hf{\frac{1}{2}}
\def\kp{\zbf k+\frac{\zbf q}{2}}
\def\km{-\zbf k+\frac{\zbf q}{2}}
\def\hwo{\hat\omega_1}
\def\hwt{\hat\omega_2}

\begin{abstract}
The in-medium properties of $\eta$ mesons are studied in hot and dense isospin asymmetric strange magnetized matter using the chiral SU(3) hadronic mean-field model. The scalar and vector density of baryons are expressed in terms of thermal distribution functions at finite temperature and magnetic field and have dependence on the scalar fields $\sigma$, $\zeta$, and $\delta$ through the effective mass of baryons and the dependence on the vector fields $\omega, \rho$, and $\phi$ through the effective chemical potential. The properties of $\eta$ mesons get modified in a hot and dense magnetized environment through the medium-modified nucleons and hyperons. The negative mass shift obtained gives rise to optical potential, which is attractive in the medium and suggests a possibility of $\eta$-mesic bound states formation. The addition of hyperons in the medium enhances the attractive interactions and causes an increased negative mass shift for $\eta$ mesons.



\end{abstract}

\maketitle

\maketitle

\section{Introduction}
\label{intro}

Exploring the in-medium properties of hadrons in the nuclear and strange medium is a fascinating research topic because of its implications in heavy-ion collision (HIC) experiments and for the physics of compact astrophysical objects. The parameters of interest as a function of which one should explore the in-medium hadron properties are: the baryonic density of the medium, temperature, isospin asymmetry, and strangeness. Various theoretical approaches to investigate the properties of matter at finite density and temperature are, for instance, coupled channel approach \cite{Tolos2004,Tolos2006,Tolos2008}, quark meson coupling (QMC) model \cite{Tsushima1999,Saito1994}, Polyakov quark Meson (PQM) model \cite{Skokov10}, Polyakov loop extended NJL (PNJL) model \cite{Blaschke12,Blaschke20}, chiral perturbation theory (ChPT) \cite{Kaplan86,Zhong2006}, QCD sum rules  \cite{Hayashigaki2000,Hilger2009,Klingl1997,Klingl1999,KMishraJpsi}, and Chiral SU(3) model \cite{Papazoglou1999,Mishra2004a,Mishra2004}. The effects of finite density and temperature on the characteristics of many mesons, such as, light vector mesons \cite{Mishra15,Mishra2019a,Pallabi22}, kaons \cite{Kaplan86,Kaplan87,Mishra2004,Mishra2009}, $D$ mesons \cite{Hayashigaki2000,Saito2007,Reddy18}, $B$ mesons \cite{Hilger2009,Dhale18,CMacho14,DP2015}, charmonium \cite{Kumar2011,Rahul17} and bottomonium \cite{DP14,Pallabi21} have been studied in literature.


Understanding how the characteristics of hadrons change in the medium when an external magnetic field is present, along with finite density and temperature, is also a current active topic of research \cite{CMacho14,Gubler16,Cho2014}. The motivation for such studies comes from the non-central HICs and the compact stars with huge magnetic fields known as magnetars. The estimated strength of the magnetic field produced in HICs is $eB \approx m_\pi^2$ at the Relativistic Heavy-Ion Collider (RHIC), and $eB \approx 15 m_\pi^2$ at the Large Hadron Collider (LHC) \cite{Deng12,Tuchin13,Skokov09}. Additionally, the magnetic field at the inside and outside of magnetars may be as strong as $10^{18}$ gauss and $10^{14}$ gauss, respectively \cite{Iwazaki,Broderick2000}. The possibility of the chiral magnetic effect is also discussed in Refs.\@ \cite{Kharzeev13,Fukushima08}, which shows that the imbalanced chirality can induce the current along the magnetic field. Moreover, the involvement of strange particles draws more attention due to its associated analysis in HICs \cite{Hartnack12, Froster07, Fuchs06}. The strange quarks are confined to respective hyperons and are supposed to hold their identification in the bound system \cite{Wang99}. For the first time, the study of the strange hadronic matter was discussed by Glendenning \cite{Glen81}, followed by Kaplan and Nelson to evaluate the in-medium properties of kaons and antikaons  \cite{Kaplan86,Kaplan87}.

The objective of present research paper is to investigate the properties of pseudoscalar $\eta$ mesons in strange hadronic matter under an external magnetic field.
The attractive nature of $\eta N$ interactions was discovered by Haider and Liu, and they proposed the possibility for $\eta$-meson bound states \cite{Haider1986,Liu1986}. The negative mass shift has prompted other researchers to investigate the possibility of $\eta N$ interactions. Considering leading-order terms, the chiral coupled channel approach predicted an optical potential of -20 MeV at nuclear saturation density ($\rho_0$) \cite{Waas}. Following the same approach and assuming the domination of $N^\ast (1535)$ over $\eta N$ interaction, the optical potential of -34 MeV was obtained in normal nuclear matter, and $\eta$ meson-bound state was anticipated with light and heavy nuclei \cite{Chiang1990}. The QMC model was used to calculate the optical potential of 60 MeV at $\rho_0$ \cite{Tsushima98}. From the chiral unitary approach of $s$-wave $\eta N$ interaction, the self-energy and optical potential of -54 MeV was obtained \cite{INOUE2002}. On taking $\eta$ self-energy as an energy density function, the result observed was $U_\eta \approx$ -72 MeV \cite{Wang2010}. Moreover, Zhong also deduced the $\eta N$ interaction from ChPT and combined Relativistic Mean Field (RMF) theory for the nucleon system. These observation leads to the in-medium $\eta$ meson mass of $(0.84 \pm 0.015)m_\eta$ and the respective optical potential of -$(83 \pm 5)$ MeV \cite{Zhong2006}. In addition, Song \textit{et al.} detected the optical potential by altering the scattering length using the RMF theory \cite{Song2008}. We can conclude that different models give different predictions for the isospin symmetric optical potential ranging from -20 MeV to -85 MeV. Along with theory, various experimental research has been conducted \cite{Peng87,Martinz99,Agakisiev13,Berg94,Chiavassa98,Averbeck}. Some of them can be listed as the study of $\eta$-meson production for various $\eta$-hadron interactions \cite{Peng87,Martinz99,Agakisiev13}, measurement of meson transverse momenta spectra in HICs around the free N-N production threshold \cite{Agakisiev13}. Moreover, the LAMPF experiments revealed that mesons are created in pion-induced nuclear processes \cite{Haider_2015}, but still, there is a need to explore hadrons at high density and temperature. However, a significant outcome toward understanding these properties can be expected from the experimental observations of BES-III \cite{BES2022}, CBM \cite{CBM22}, J-PARC \cite{JPARC2022}, NICA \cite{NICA2022}, and PANDA \cite{GSI2022}. 


%

In the present work, we have utilized a non-perturbative RMF model known as the chiral SU(3) model to investigate the $\eta$-baryon interactions in the strange hadronic medium. The model is based on the non-linear realization of the chiral SU(3)$_L$ × SU(3)$_R$ symmetry, and it respects the QCD fundamental properties such as chiral symmetry breaking (spontaneous and explicit) and broken scale invariance \cite{ZschPhy00}. In the model, all mesonic fields are treated as classical fields, resulting in the zero expectation value of mesons except vector and scalar meson fields which add to the nucleon–meson interaction Lagrangian term \cite{Mishra2004a}. The model successfully describes the finite nuclei \cite{Wang2002,Gam}, hypernuclei \cite{Tsub2010}, strange hadronic matter \cite{WANG2001,Harpreet18,Rahul2015,Kumar2020phi,Rahul2018,Kumar2011,DP2015}, nuclear matter \cite{Kumar2019,Kumar20D,KMishraJpsi,KKJpsi19,Rajesh2020eta}, and neutron stars \cite{KMishraNstar,Shivam19, Kumari2022}. The optical potentials of $\eta$ mesons in non-magnetized and magnetized nuclear mediums have been studied in refs. \cite{Rajesh2020eta, Kumar2022etam}. As the matter produced in the HICs can also be strange, it is imperative to study $\eta$ meson interactions by considering the full baryon octet. To our knowledge, no work in the literature has explored properties of $\eta$ mesons in a magnetized strange medium (having both nucleons and hyperons). 


The outline of this paper is as follows: in \cref{Sec. II} the details of the chiral model are given along with the in-medium dispersion relations for $\eta$ meson. Particularly in \cref{Sec. IIA 1}, we have derived the thermodynamic potential per unit volume for charged baryons in the magnetic field. Whereas, in \cref{Sec. IIA 2}, a similar description for uncharged baryons is provided. In \cref{Sec. IIB}, we have derived the $\mathcal{\eta B}$ interactions in the chiral model. In \cref{Sec. III}, the analysis for fields, scalar densities of baryons, mass modification, and optical potential of $\eta$ meson is presented. Finally, the entire paper winds up to a summary.

\section{Methodology}
\label{Sec. II}
\subsection{In-medium Interactions in the Chiral Model}
\label{Sec. IIA}
For describing the hadron-hadron interactions, we have used the approach of effective mean field theory \cite{Papazoglou1999,Mishra2004,Mishra2004a}, which is based on the non-linear realization of chiral symmetry \cite{Veronica12,Zsch2000,Weinberg68,Bardeen69} at finite temperature and density along with the presence of the external magnetic field. The Lagrangian density of the chiral SU(3) model is defined as
\be
{\cal L}_{\text{chiral}} = {\cal L}_{kin} + \sum_{ M =S,V}{\cal L}_{BM}
+ {\cal L}_{vec} + {\cal L}_0 + {\cal L}_{SB}.
\label{genlag} \ee

The first term, ${\cal L}_{kin}$ in \cref{genlag}, defines the kinetic energy, and the next ${\cal L}_{BM}$ term describes the interaction between mesons (having spin-0 and spin-1 mesons as S and V respectively) and baryons, given by  
\begin{eqnarray}
	{\cal L}_{BM} = - \sum_{i} \bar {\psi_i} 
	\left[ m_{i}^{*} + g_{\omega i} \gamma_{0} \omega 
	+ g_{\rho i} \gamma_{0} \tau_3 \rho + g_{\phi i} \gamma_{0} \phi \right] \psi_{i},
	\label{BM}
\end{eqnarray}
where ($i =p, n, \Lambda, \Sigma^+,\Sigma^0,\Sigma^-,\Xi^-,\Xi^0$) and the in-medium baryon mass $m_{i}^{*}$ is given as
\begin{eqnarray}
	m_{i}^{*} = -(g_{\sigma i}\sigma + g_{\zeta i}\zeta + g_{\delta i}\tau_3 \delta),
	\label{massn}
\end{eqnarray}
here $\tau_3$ stands for the isospin third component, and the coupling constants of  $\sigma$,  $\zeta$, and $\delta$ fields are given by $g_{\sigma i}$, $g_{\zeta i}$ and $g_{\delta i}$ respectively. 
The third, self-interaction term $ {\cal L}_{vec}$ for mass reproduction of vector meson is given by the Lagrangian,
\begin{eqnarray}
	{\cal L} _{vec} & = & \frac {1}{2} \left( m_{\omega}^{2} \omega^{2} 
	+ m_{\rho}^{2} \rho^{2} + m_{\phi}^{2} \phi^{2}\right) 
	\frac {\chi^{2}}{\chi_{0}^{2}}
	+  g_4 (\omega ^4 +6\omega^2 \rho^2+\rho^4+2\phi^4).
	\label{vec}
\end{eqnarray}
The spontaneous chiral symmetry breaking or the self-interaction term of scalar mesons, ${\cal L}_{0}$ is given by 
\begin{align}
	{\cal L} _{0}  =  -\frac{1}{2} k_{0}\chi^{2} \left( \sigma^{2} + \delta^{2} +\zeta^{2} 
	\right) + k_{1} \left( \sigma^{2} + \delta^{2} +\zeta^{2} 
	\right)^{2} \nonumber\\
	+ k_{2} \left( \frac {\sigma^{4}}{2} + \frac {\delta^{4}}{2} + 3 \sigma^{2} 
	\delta^{2} + \zeta^{4} \right) 
	+ k_{3}\chi\zeta\left( \sigma^{2} - \delta^{2} \right) \nonumber\\
	- k_{4} \chi^{4} 
	-  \frac {1}{4} \chi^{4} {\rm {ln}} 
	\frac{\chi^{4}}{\chi_{0}^{4}}
	+ \frac {d}{3} \chi^{4} {\rm {ln}} \Bigg (\bigg( \frac {\left( \sigma^{2} 
		- \delta^{2}\right) \zeta }{\sigma_{0}^{2} \zeta_{0}} \bigg) 
	\bigg (\frac {\chi}{\chi_0}\bigg)^3 \Bigg ),
	\label{lagscal}
\end{align}
where $\sigma_0$, $\delta_0$, $\zeta_0$, and $\chi_0$ represent the respective vacuum values of the fields, $\sigma$, $\delta$, $\zeta$, and $\chi$. 

The explicit chiral symmetry breaking term, ${\cal L}_{SB} $ is given by
\begin{align}
	{\cal L} _{SB} =  -\left( \frac {\chi}{\chi_{0}}\right)^{2} 
	\left[ m_{\pi}^{2} 
	f_{\pi} \sigma
	+ \big( \sqrt {2} m_{K}^{2}f_{K} - \frac {1}{\sqrt {2}} 
	m_{\pi}^{2} f_{\pi} \big) \zeta \right],
	\label{lsb}
\end{align} 
where $f_\pi$, $f_K$, $m_\pi$, $m_K$ represent the decay constants and masses of pions and kaons, whose values are given in \cref{tab}. 
The partition function of grand canonical ensemble for the hadronic system is given by \cite{KKJpsi19,Zschiesche1997} 
\bea
\mathcal{Z} &=& \Tr \exp[-\beta(\hat{\cal{H}}-\sum_{i}
\mu_i \hat{\mathcal{N}}_i)],
\label{pfunc}
\eea
where $\beta=1/T$ and $\hat{\mathcal{H}}$, $\hat{\mathcal{N}}_i$ are the operators of Hamiltonian density and number density, respectively whereas $\mu_i $ represents the chemical potential. At given temperature $T$, the thermodynamic potential, $\Omega$ is defined as
\bea
\Omega(T,V,\mu)=-T \ln \cal{Z}.
\eea
Under the mean-field approximation, thermodynamic potential per unit volume, $\frac{\Omega} {V}$, at zero magnetic field, in chiral SU(3) model transforms into the following equation,
\begin{equation}
\label{thermo}
\frac{\Omega} {V}= -\frac{\gamma_i T}
{(2\pi)^3} \sum_{i}
\int d^3{k}\biggl\{{\rm ln}
\left( 1+e^{-\beta [ E^{\ast}_i(k) - \mu^{*}_{i}]}\right) \\
+ {\rm ln}\left( 1+e^{-\beta [ E^{\ast}_i(k)+\mu^{*}_{i} ]}
\right) \biggr\} -{\cal L}_{vec} - {\cal L}_0 - {\cal L}_{SB}-{\mathcal{V}}_{vac}.  
\end{equation}
In the above expression, $\gamma_i$, and 
$ \mu^{*}_{i}=\mu_{i}-g_{\omega i}\omega-g_{\rho i}\tau_{3}\rho$ are the spin degeneracy factor, and effective chemical potential of baryons respectively; whereas  $E^{\ast}_i(k)=\sqrt{k^2+m^{*}_{i}}$ represents the effective single particle energy with $k$ being the particle momentum in the direction of magnetic field \cite{Vero21}. For obtaining zero vacuum energy, the vacuum potential energy, ${\mathcal{V}}_{vac}$, is subtracted in \cref{thermo}.
In the presence of exeternal magnetic field \cref{genlag} modifies to
\begin{equation}
{\cal L}_{T}={\cal L}_{chiral}+{\cal L}_{mag},
\label{Tlag}
\end{equation}
where
\be 
{\cal L}_{mag}=-{\bar {\psi_i}}q_i 
\gamma_\mu A^\mu \psi_i
-\frac {1}{4} \kappa_i \mu_N {\bar {\psi_i}} \sigma ^{\mu \nu}F_{\mu \nu}
\psi_i
-\frac{1}{4} F^{\mu \nu} F_{\mu \nu}.
\label{lmag}
\ee

Here, $\psi_i$ is the wave function of $i^{th}$ baryon, and the second term is the tensorial interaction with the electromagnetic field tensor, $F_{\mu \nu}$. The second tensorial interaction term of the above equation is related to the anomalous magnetic moment (AMM) of  $i^{th}$ baryon. It contains the term $\mu_N$ ($={e}/{2m_N}$), which refers to the nuclear Bohr magneton having $m_N$ as the nucleon vacuum mass. As we have considered the uniform magnetic field ($B$) along the $Z$-axis, so the value of vector potential will be $A^\mu =(0,0,Bx,0)$. 

\subsubsection{Charged Baryons in Magnetic Field}
\label{Sec. IIA 1}
We must consider Lorentz force due to the charged baryons for a uniform magnetic field. Therefore, the momenta of charged baryons are divided into two components: transverse momenta, $k_\perp^{b}$, and longitudinal momenta, $k_\parallel^{b}$. The $k_\perp^{b}$ with an electric charge $q_b$ are confined to discrete Landau levels, $\nu$, with, $(k_\perp^{b})^2 = 2 \nu |q_b| B$, where $\nu \geq 0$ is an integral quantum number \cite{Strickland2012}. Hence, the expression for conversion of volume integral for magnetic field calculations is

\begin{equation}
\int {d^3}k \rightarrow \frac{|q_b| {B}}{(2\pi)^2} 
\sum_n \int_{0}^\infty d k_\parallel^{b} \, ,
\label{lsumform}
\end{equation}
where the summation $n$ is over the discrete orbital angular momentum of baryons in the perpendicular plane, and the momenta $k_\parallel^{b}$ is taken along the direction of the magnetic field. The orbital quantum number relates to $\nu$ by the relation
$\nu=n+\frac{1}{2}-\frac{q_b}{|q_b|}\frac{s}{2}=0, 1, 2, \ldots$. Here, $s=\pm 1$ denotes the spin projection of the particle along the magnetic field \cite{Vero21}. Also, the effective single particle baryon energy will be quantized \cite{KKJpsi19} and given as
\bea
\tilde E^{b}_{\nu, s}&=&\sqrt{\left(k^{b}_{\parallel}\right)^{2}+
\left(\sqrt{m^{* 2}_{b}+2\nu |q_b|{B}}-s\mu_{N}\kappa_{b}{B}\right)^{2}}.
\eea

Therefore, the first term of \cref{thermo} for charged baryon will be
\bea
\label{thermop}
&\frac{\Omega^{C}_b} {V}= -
\frac{ T|q_b| B}{(2\pi)^2} \Bigg[
\sum_{\nu=0}^{\nu_{max}^{(s=1)}} \int_{0}^\infty d k_\parallel^{b} \, \biggl\{{\rm ln}
\left( 1+e^{-\beta [ \tilde E^{b}_{\nu, s} - \mu^{*}_{b} ]}\right) 
+ {\rm ln}\left( 1+e^{-\beta [ \tilde E^{b}_{\nu, s}+\mu^{*}_{b} ]}
\right) \biggr\}\nonumber\\
&+
\sum_{\nu=1}^{\nu_{max}^{(s=-1)}} \int_{0}^\infty d k_\parallel^{b} \, \biggl\{{\rm ln}
\left( 1+e^{-\beta [ \tilde E^{b}_{\nu, s} - \mu^{*}_{b} ]}\right)+ {\rm ln}\left( 1+e^{-\beta [ \tilde E^{b}_{\nu, s}+\mu^{*}_{b} ]}
\right) \biggr\}\Bigg].  
\eea

\subsubsection{Uncharged Baryons in Magnetic Field}
\label{Sec. IIA 2}
Landau quantization is not present in the case of uncharged baryons with the external magnetic field (as it is the quantization of the cyclotron orbits of charged particles in a uniform magnetic field and only charged particles can occupy Landau levels), so there is no change in the term $\int {d^3k}/{(2\pi)^3} \,$ \cite{Strickland2012}. However, due to the non-zero value of AMM of uncharged baryons, the thermodynamic potential of \cref{thermo} will be modified as
\begin{equation}
\label{thermon}
\frac{\Omega^{U}_{b}} {V}= -\frac{T}
{(2\pi)^3} \sum_{s=\pm 1}
\int d^3k\biggl\{{\rm ln}
\left( 1+e^{-\beta [ \tilde E^{b}_{s} - \mu^{*}_{b} ]}\right) \\
+ {\rm ln}\left( 1+e^{-\beta [\tilde E^{b}_{s}+\mu^{*}_{b}]}
\right) \biggr\},   
\end{equation}
where $\tilde E^{b}_{s}$ is the effective single particle energy of uncharged baryons in the presence of  magnetic field and is given by
\bea
\tilde E^{b}_{s}&=& \sqrt{\left(k^{b}_{\parallel}\right)^{2} +
\left(\sqrt{m^{* 2}_{b}+\left(k^{b}_{\bot}\right)^{2} }-s\mu_{N}\kappa_{b}{B}
\right)^{2}}.
\eea

In the presence of external magnetic field, the net thermodynamical potential is written as
\begin{equation}
\label{thermonet}
\frac{\Omega} {V}= \frac{\Omega^{C}_b} {V}+\frac{\Omega^{U}_b} {V} -{\cal L}_{vec} - {\cal L}_0 - {\cal L}_{SB}-{\mathcal{V}}_{vac}. 
\end{equation}

When we minimize the thermodynamical potential density, $\Omega/V$ of the strange system, then the resulting coupled equations of the mesonic fields $\sigma$, $\zeta$, $\delta$, $\omega$, $\rho$, $\phi$, and $\chi$, are determined as 
\begin{align}
\frac{\partial (\Omega/V)}{\partial \sigma}= k_{0}\chi^{2}\sigma-4k_{1}\left( \sigma^{2}+\zeta^{2}
+\delta^{2}\right)\sigma-2k_{2}\left( \sigma^{3}+3\sigma\delta^{2}\right)
-2k_{3}\chi\sigma\zeta \nonumber\\
-\frac{d}{3} \chi^{4} \bigg (\frac{2\sigma}{\sigma^{2}-\delta^{2}}\bigg )
+\left( \frac{\chi}{\chi_{0}}\right) ^{2}m_{\pi}^{2}f_{\pi}
-\sum g_{\sigma i}\rho_{i}^{s} = 0,
\label{sigma}
\end{align}
\begin{align}
\frac{\partial (\Omega/V)}{\partial \zeta}= k_{0}\chi^{2}\zeta-4k_{1}\left( \sigma^{2}+\zeta^{2}+\delta^{2}\right)
\zeta-4k_{2}\zeta^{3}-k_{3}\chi\left( \sigma^{2}-\delta^{2}\right)\nonumber\\
-\frac{d}{3}\frac{\chi^{4}}{\zeta}+\left(\frac{\chi}{\chi_{0}} \right)
^{2}\left[ \sqrt{2}m_{K}^{2}f_{K}-\frac{1}{\sqrt{2}} m_{\pi}^{2}f_{\pi}\right]
 -\sum g_{\zeta i}\rho_{i}^{s} = 0 ,
\label{zeta}
\end{align}
\begin{align}
\frac{\partial (\Omega/V)}{\partial \delta}=k_{0}\chi^{2}\delta-4k_{1}\left( \sigma^{2}+\zeta^{2}+\delta^{2}\right)
\delta-2k_{2}\left( \delta^{3}+3\sigma^{2}\delta\right) +2k_{3}\chi\delta
\zeta \nonumber\\
 +   \frac{2}{3} d \chi^4 \left( \frac{\delta}{\sigma^{2}-\delta^{2}}\right)
-\sum g_{\delta i}\tau_3\rho_{i}^{s} = 0 ,
\label{delta}
\end{align}
\begin{align}
\frac{\partial (\Omega/V)}{\partial \omega}=\left (\frac{\chi}{\chi_{0}}\right) ^{2}m_{\omega}^{2}\omega+g_{4}\left(4{\omega}^{3}+12{\rho}^2{\omega}\right)-\sum g_{\omega i}\rho_{i}^{v} = 0 ,
\label{omega}
\end{align}
\begin{align}
\frac{\partial (\Omega/V)}{\partial \rho}=\left (\frac{\chi}{\chi_{0}}\right) ^{2}m_{\rho}^{2}\rho+g_{4}\left(4{\rho}^{3}+12{\omega}^2{\rho}\right)-\sum g_{\rho i}\tau_3\rho_{i}^{v} = 0 ,
\label{rho}
\end{align}
\begin{align}
\frac{\partial (\Omega/V)}{\partial \phi}=\left (\frac{\chi}{\chi_{0}}\right) ^{2}m_{\phi}^{2}\phi+ 8 g_{4}{\phi}^{3}-\sum g_{\phi i}\rho_{i}^{v} = 0 ,
\label{phi}
\end{align}
and
\begin{align}
\frac{\partial (\Omega/V)}{\partial \chi}=k_{0}\chi \left( \sigma^{2}+\zeta^{2}+\delta^{2}\right)-k_{3}
\left( \sigma^{2}-\delta^{2}\right)\zeta + \chi^{3}\left[1
+{\rm {ln}}\left( \frac{\chi^{4}}{\chi_{0}^{4}}\right)  \right]
+(4k_{4}-d)\chi^{3}
\nonumber\\
-\frac{4}{3} d \chi^{3} {\rm {ln}} \Bigg ( \bigg (\frac{\left( \sigma^{2}
-\delta^{2}\right) \zeta}{\sigma_{0}^{2}\zeta_{0}} \bigg )
\bigg (\frac{\chi}{\chi_0}\bigg)^3 \Bigg )+
\frac{2\chi}{\chi_{0}^{2}}\left[ m_{\pi}^{2}
f_{\pi}\sigma +\left(\sqrt{2}m_{K}^{2}f_{K}-\frac{1}{\sqrt{2}}
m_{\pi}^{2}f_{\pi} \right) \zeta\right] \nonumber\\
- \frac{\chi}{{\chi_0}^2}(m_{\omega}^{2} \omega^2+m_{\rho}^{2}\rho^2+m_{\phi}^{2}\phi^2)  = 0 ,
\label{chi}
\end{align}
respectively.
The parameters $k_0, k_2$ and $k_4$ are introduced to regenerate the vacuum mass of $\sigma$, $\zeta$, and $\chi$ meson, whereas the constant  $k_1$, is to produce the effective nucleon mass at saturation density around 0.$65 m_N$ and $k_3$ is the constraint by $\eta$ and $\eta^\prime$ masses \cite{Zschiesche}. To get the exact nuclear properties (${E}/{\rho_B} - m_N$ = -16 MeV, $E_{sym}$=32 MeV, $K=250.6$ MeV at $\rho_0=0.15$ fm$^{-3}$) we have adjusted $\chi_0$ and $g_4$. Also, in the above equations, $\rho^{s}_{i}$ and $\rho^{v}_{i}$ represent the scalar and vector or number densities of $i^{th}$ baryon, respectively.

The vector and scalar densities of charged baryons in the presence of magnetic field are written as \cite{Broderick2000}

\begin{align}
\rho^{v}_{b}=\frac{|q_{b}|{B}}{2\pi^2} \Bigg [ 
\sum_{\nu=0}^{\nu_{max}^{(s=1)}} \int^{\infty}_{0}
dk^b_{\parallel}\left( f^b_{k,\nu, s}-\bar{f}^b_{k,\nu, s}\right)
+\sum_{\nu=1}^{\nu_{max}^{(s=-1)}} \int^{\infty}_{0}
dk^b_{\parallel}\left( f^b_{k,\nu, s}-\bar{f}^b_{k,\nu, s}\right) 
\Bigg],
\label{rhovp}
\end{align}
and
\begin{equation}
\rho^{s}_{b}=\frac{|q_{b}| {B}m^{*}_{b}}{2\pi^2} \Bigg [ 
\sum_{\nu=0}^{\nu_{max}^{(s=1)}}\int^{\infty}_{0}\frac{dk^b_{\parallel}}{\sqrt{(k^{b}_{\parallel})^2
+(\bar m_{b})^2}}\left( f^b_{k,\nu, s}+\bar{f}^b_{k, \nu, s}\right)
+\sum_{\nu=1}^{\nu_{max}^{(s=-1)}} \int^{\infty}_{0}\frac{dk^b_{\parallel}}{\sqrt{(k^{b}_{\parallel})^2
+(\bar m_{b})^2}}\left( f^b_{k,\nu, s}+\bar{f}^b_{k, \nu, s}\right)
\Bigg],
\label{rhosp}
\end{equation}
respectively, where $\bar m_{b}$ denotes the effective mass under the effect of the magnetic field, which is given as
\begin{align}
\bar m_{b}=\sqrt{m^{* 2}_{b}+2\nu |q_{b}|{B}}-s\mu_{N}\kappa_{b}{B}.
\label{mc}
\end{align}

Similarly, the number and scalar densities for the uncharged baryons will be \cite{Broderick2000},
\begin{align}
\rho^{v}_{b}=\frac{1}{2\pi^{2}}\sum_{s=\pm 1}\int^{\infty}_{0}k^{b}_{\bot}
dk^{b}_{\bot} \int^{\infty}_{0}\, dk^{b}_{\parallel}
\left( f^b_{k, s}-\bar{f}^b_{k, s}\right), 
\label{rhovn} 
\end{align}
and
\begin{align}
\rho^{s}_{b}=\frac{1}{2\pi^{2}}\sum_{s=\pm 1}\int^{\infty}_{0}
k^{b}_{\bot}dk^{b}_{\bot}\left(1-\frac{s\mu_{N}\kappa_{b}{B}}
{\sqrt{m^{* 2}_{b}+\left(k^{b}_{\bot}\right)^{2}}} \right)  
\int^{\infty}_{0}\, dk^{b}_{\parallel}
\frac{m^*_b}{\tilde E^{b}_{s}}\left(
f^b_{k, s}+\bar{f}^b_{k, s}\right),
\label{rhosn} 
\end{align}
respectively. In the above equations, ${f}^b_{k, \nu, s}$, $\bar{f}^b_{k, \nu, s}$,  ${f}^b_{k, s}$ and $\bar{f}^b_{k, s}$ represent the finite temperature distribution functions for particles and antiparticles for the respective charged and uncharged baryons, and given as
\begin{align}
f^b_{k,\nu, s} = \frac{1}{1+\exp\left[\beta(\tilde E^b_{\nu, s} 
-\mu^{*}_{b}) \right]}, \qquad
\bar{f}^b_{k,\nu, s} = \frac{1}{1+\exp\left[\beta(\tilde E^b_{\nu, s} 
+\mu^{*}_{b} )\right]},
\label{dfp}
\end{align}
\begin{align}
f^b_{k, s} = \frac{1}{1+\exp\left[\beta(\tilde E^b_{s} 
-\mu^{*}_{b}) \right]}, \qquad
\bar{f}^b_{k, s} = \frac{1}{1+\exp\left[\beta(\tilde E^b_{s} 
+\mu^{*}_{b} )\right]}.
\label{dfn}
\end{align}

\subsection{$\cal{\eta B}$ interactions in the Chiral Model}
\label{Sec. IIB}
The main motive of the present work is to study the interactions of $\eta$ mesons with baryon octet. Within the chiral SU(3) model, the interaction Lagrangian density for $\eta-B$ system is given by
	\begin{align} \label{etaN}
		\mathcal{L_{\eta B}}  =
		\left( \frac{1}{2}-\frac{\sqrt{2}\sigma ^\prime f_\pi + 4 \zeta ^\prime (2 f_K-f_\pi) }{\sqrt{2}f^2} \right) \partial^{\mu}\eta\partial_{\mu}\eta 
		-\frac{1}{2}\left(
		m_{\eta}^2
		-\frac{(\sqrt{2}\sigma ^\prime -4 \zeta ^\prime )m^2_\pi f_\pi + 8 \zeta ^\prime m^2_K f_K}{\sqrt{2} f^2}
		\right) \eta^2\nonumber\\
		+\frac{\partial^{\mu}\eta\partial_{\mu}\eta}{4 f^{2}}\left(d^{'} \Sigma \rho_{i}^{s}+2 d_{2} \rho_{\Lambda^{0}}^{s}+3 d_{2}\left(\rho_{\Xi^-}^{s}+\rho_{\Xi^0}^{s}\right)\right).
	\end{align}
 In the following, we describe how the different terms of above Lagrangian density are obtained.
	\subsubsection{Range term}
	The first term of \cref{etaN} is known as the range term
 and is obtained from the Lagrangian density   \cite{Papazoglou1999} 
	\begin{equation}
		\label{pikin}
		{\mathcal L}_{{\mathrm{1st range term}}} =  Tr (u_{\mu} X u^{\mu}X +X u_{\mu} u^{\mu} X).
	\end{equation}
	
	The term $u_{\mu}$ is defined as: $u_{\mu}=-\frac{i}{2}\left[u^{\dagger}(\partial_{\mu}u) -u(\partial_{\mu}u^\dagger) \right]$ and
	$u$=$\text{exp}\left[\frac{i}{\sqrt{2}\sigma_{0}}P\gamma_{5}\right]$. These are expanded up to second order to obtain the detailed expression. Symbols $X$ and $P$ denote scalar and pseudoscalar meson matrices.
 We have
	
	The vacuum values of the fields, $\sigma$, and $\zeta$ can be calculated using axial currents of kaons and pions by these relations \cite{1998Papazoglou},
	\be
	\label{zeta0}
	\sigma_0 = -f_{\pi} \qquad \zeta_0 = -\frac{1}{\sqrt{2}}(2 f_K - f_{\pi}).
	\ee 
	The terms $\sigma'$, $\zeta'$, and $\delta'$ represent the digression of field values from their expectation values to the vacuum expectations, 
	\be
	\sigma'=(\sigma-\sigma_0), \qquad
	\zeta'=(\zeta-\zeta_0), \qquad
	\delta'=(\delta-\delta_0).
	\ee
	\subsubsection{Mass term}
	The second line of \cref{etaN}, known as mass term in the $\mathcal{\eta B}$ Lagrangian density is derived from 
	\begin{equation}
		\label{esb-gl}
		\mathcal{L}^{\eta}_{S\mathcal {B}}=
		-\frac{1}{2} \Tr A_p \left(uXu+u^{\dagger}Xu^{\dagger}\right),
	\end{equation}	
	where $A_p$ represents a diagonal matrix given as
	\begin{equation}
		\label{apmat}
		A_p=\frac{1}{\sqrt{2}}
		\left( \begin{array}{ccc}
			m_{\pi}^2 f_{\pi}& 0& 0\\   
			0 & m_\pi^2 f_\pi& 0\\
			0 & 0& 2 m_K^2 f_K
			-m_{\pi}^2 f_\pi
		\end{array} \right).
	\end{equation}
	The vacuum mass of $\eta$ meson, $m_{\eta}$, deduced from \cref{esb-gl} can be written as
	\begin{equation}
		m_{\eta}=\frac{1}{f}\sqrt{\frac{1}{2}(-\left(8 f_K f_\pi (m_\pi^2+m_K^2))+16 f_K^2 m_K^2 +6 f_\pi^2 m_\pi^2\right)},
	\end{equation}
 where $f = \sqrt{f_\pi^2 +2 \left(2f_K - f_\pi\right)^2}$.
	\subsubsection{$d$ terms}
	The $d$ terms (last line in \cref{etaN}) are also known as range terms and are obtained from baryon-meson interaction Lagrangian densities \cite{Mishra2006, Mishra2004}	
	\begin{equation}
		{\cal L }_{d_1}^{\bar{\mathcal{B}} \mathcal{B}} =\frac {d_1}{2} \Tr (u_\mu u ^\mu) \Tr(\bar{\mathcal{B}} \mathcal{B}),
		\label{L1}
	\end{equation}
	and
	\begin{equation}
		{\cal L }_{d_2}^{\bar{\mathcal{B}} \mathcal{B}} =d_2 \Tr (\bar {\mathcal{B}} u_\mu u ^\mu \mathcal{B}),
		\label{L2}
	\end{equation}
where ${\mathcal{B}}$ is the baryon matrix given as \cite{Rajesh2020eta}	
	\be
	\label{bmat}
	\mathcal{B}=\frac{1}{\sqrt{2}}b^a \lambda_a=
	\left( \begin{array}{ccc}
		\frac{\Sigma^0}{\sqrt{2}} +\frac{\Lambda}{\sqrt{6}}& \Sigma^+ & p\\   
		\Sigma^- & -\frac{\Sigma^0}{\sqrt{2}} +\frac{\Lambda}{\sqrt{6}} & n \\
		\Xi^- & \Xi^0& -2 \frac{\Lambda}{\sqrt{6}}
	\end{array} \right).
	\ee
	On further solving \cref{L1} and \cref{L2} we get the following equations	
    \begin{equation}
{\cal L}_{d_{1}}^{\bar{\mathcal{B}} \mathcal{B}}=\frac{3 d_{1}}{4 \sigma_{0}^{2} c^{2}} {\partial^{\mu}\eta\partial_{\mu}\eta}\left[\rho^s_{p}+\rho^s_{n}+\rho^s_{\Sigma^{+}}+\rho^s_{\Sigma^{0}}+\rho^s_{\Sigma^{-}}+\rho^s_{\Lambda^{0}}+\rho^s_{\Xi^0}+\rho^s_{\Xi^-}\right],
    \end{equation}
	and
	\begin{equation}
{\cal L}_{d_{2}}^{\bar{\mathcal{B}} \mathcal{B}}=\frac{d_{2}}{4 \sigma_{0}^{2} c^{2}}{\partial^{\mu}\eta\partial_{\mu}\eta}\left[\rho^s_{\Sigma^{-}}+\rho^s_{\Sigma^{0}}+\rho^s_{\Sigma^{+}}+\rho^s_{p}+\rho^s_{n}+4\left(\rho^s_{\Xi^-}+\rho^s_{\Xi^0}\right)+3 \rho^s_{\Lambda^{0}}\right].
\end{equation}\\
Here $c= {2 \sigma_0}/{(\sqrt{2} \zeta_0 + \sigma_0)}$.
Hence, the resulting $d$ term is given as
	\begin{equation}
\frac{\partial^{\mu}\eta\partial_{\mu}\eta}{4 f^{2}}\left[d^{'} \Sigma \rho_{i}^{s}+2 d_{2} \rho_{\Lambda^{0}}^{s}+3 d_{2}\left(\rho_{\Xi^-}^{s}+\rho_{\Xi^0}^{s}\right)\right],
\end{equation}
where $d'$ = $3d_1+d_2$.
The equation of motion using $\eta \mathcal{B}$ Lagrangian density can be written as
\begin{eqnarray} 
		&& \partial^{\mu}\partial_{\mu} \eta + \left(
		m_{\eta}^2-\frac{(\sqrt{2}\sigma ^\prime -4 \zeta ^\prime )m^2_\pi f_\pi + 8 \zeta ^\prime m^2_K f_K}{\sqrt{2} f^2}
		\right)\eta  \nonumber\\
		&&+\frac{2}{f^{2}}\left(\frac{d^{\prime}  \Sigma \rho_{i}^{s}}{4}+\frac{d_{2} \rho_{\Lambda^{0}}^{s}}{2}+\frac{3d_{2} (\rho_{\Xi^-}^{s}+\rho_{\Xi^0}^{s})}{4}-\frac{\sqrt{2} \sigma^{\prime} f_{\pi}+4 \zeta^{\prime}\left(2 f_{K}-f_{\pi}\right)}{\sqrt{2}}\right) \partial^{\mu} \partial_{\mu} \eta=0.
		\label{eom}
	\end{eqnarray}\\
The dispersion relation for $\eta$ meson can be obtained by the Fourier transform of the above equation, $i.e.$,
	\begin{equation}
		-\omega^2+ { \textbf{k}}^2 + m_\eta^2 -\Pi^*(\omega, | \textbf{k}|)=0,
		\label{drk}
	\end{equation}	
	where,  $\Pi^*$ stands for $\eta$ meson's effective self-energy, which is described as
	\begin{align}
		\Pi^* (\omega, | \textbf{k}|) = 
		&&\frac{ 8 \zeta ^\prime m^2_K f_K + (\sqrt{2}\sigma ^\prime -4 \zeta ^\prime )m^2_\pi f_\pi}{\sqrt{2} f^2}
		+\frac{2}{f^{2}}\left(\frac{d^{\prime} \rho_{b}^{s}}{4}+\frac{d_{2} \rho_{\Lambda^{0}}^{s}}{2}+\frac{3d_{2} (\rho_{\Xi^-}^{s}+\rho_{\Xi^0}^{s})}{4} \right)
		(\omega ^2 - {\textbf{ k}}^2) \nonumber\\
		&&- \frac{2}{f^2} \left( \frac{\sqrt{2} \sigma ^\prime f_\pi + 4 \zeta ^\prime (2 f_K-f_\pi) }{\sqrt{2}} \right)
		(\omega ^2 - {\textbf{ k}}^2). 
		\label{sen}
	\end{align}
	
The parameter $d'$ is fitted to the values of scattering length, $a^{\eta N}$ \cite{Zhong2006}. For the present model, the scattering length expression is obtained as
\begin{align}
		a^{\eta N} = \frac{1}{4 \pi \left (1+\frac{m_\eta}{M_N}\right )} \Big [ \Big( \frac{d'}{\sqrt{2}}-\frac{g_{\sigma N}{f_\pi}}{m^2_\sigma}+\frac{4 (2f_K-f_\pi) g_{\zeta N}}{m^2_\zeta} \Big) \frac {m_\eta ^2} {\sqrt{2}f^2} \nonumber \\
		+ \left( \frac{\sqrt{2} g_{\sigma N}}{m^2_\sigma}-\frac{4 g_{\zeta N}}{m^2_\zeta} \right )\frac {m^2_\pi f_\pi} {2\sqrt{2}f^2}+ \frac{2\sqrt{2} g_{\delta N}}{m^2_\delta} \frac {m^2_K f_K} {f^2}  \Big ].
		\label{sl}
	\end{align}
	
	
  On substituting  the condition $m_{\eta}^*=\omega(| \textbf{k}|$=0) in \cref{drk}, the effective mass of $\eta$ meson in strange matter is calculated. We can also define the momentum-dependent optical potentials from the following relation \cite{Mishra2008,Mishra2009} 
	\begin{equation}
		U^*_{\eta}(\omega,\textbf{ k}) = \omega (\textbf{k}) -\sqrt {\textbf{k}^2 + m^{^2}_{\eta}}.
		\label{opk}
	\end{equation}
	
\section{Results and Discussions}
\label{Sec. III}
This section presents the results on in-medium masses and optical potential of $\eta$ mesons, which are calculated from medium-modified scalar and vector fields in magnetized asymmetric strange hadronic matter. We divide the discussion into two parts: (\cref{rs III A}) the behavior of scalar fields ($\sigma$, $\zeta$, $\delta$ and $\chi$) and scalar density ($\rho^{s}_{i}$) of baryons, (\cref{rs III B}) the medium-modified mass and optical potential of $\eta$ meson using chiral SU(3) model. Various parameters used in the present calculations are listed in \cref{tab}.
\begin{table}[H]
	\centering
	\begin{tabular}{|c|c|c|c|c|c|c|}	
		\hline
		$k_0$ & $\sigma_0$ (MeV) & $g_{\sigma N}$ & $g_{\sigma \Lambda}$ & $g_{\sigma \Sigma}$ & $g_{\sigma \Xi}$ & $m_{\sigma}$ (MeV)  \\ 
		\hline 
		2.37 & -93.3 & 9.83 & 6.734 & 5.22 & 2.882 & 466.5  \\ 

		\hline 
		$k_1$& $\zeta_0$ (MeV) & $g_{\zeta N }$  & $g_{\zeta \Lambda}$  & $g_{\zeta \Sigma}$  &$g_{\zeta \Xi}$  & $m_{\zeta}$ (MeV)  \\ 
		\hline 
		1.40 & -106.763 & -1.22 & 3.158 & 5.296 & 8.605 & 1024.5  \\

		\hline
		$k_2$  & $\delta_0$ (MeV)   &  $g_{\delta N }$  & $g_{\delta \Lambda}$  & $g_{\delta \Sigma}$  & $g_{\delta \Xi}$ & $m_{\delta}$  (MeV) \\

		\hline 
		-5.55 & 0 & 2.34 & 0 & 3.473 & 2.340 & 899.5  \\

		\hline 
		$k_3$ & $\chi_0$ &$g_{\omega N}$  & $g_4$ & $f_{\pi}$ (MeV) & $m_{\pi}$ (MeV) & $M_N$ (MeV) \\ 
		\hline 
		-2.65 & 401.91 & 12.13 & 38.9 & 93.3 & 139 & 939  \\ 
		\hline
		
		$k_4$ & $\rho_0$ ($\text{fm}^{-3}$)  &  $g_{\rho N}$ & $d$ & $f_K$ (MeV) & $m_K$ (MeV) & $m_{\eta}$ (MeV) \\
		\hline
		
		-0.232 & 0.15 & 4.92 & 0.06 & 122.143 & 498 & 574.395 \\ 
		\hline	
	\end{tabular}
	\caption{Parameters used in the present work} \label{tab}
\end{table}

\subsection{In-medium behavior of scalar mean fields and densities} 
\label{rs III A}
\begin{figure}[ht!] 
	\includegraphics[width=16cm,height=22cm]{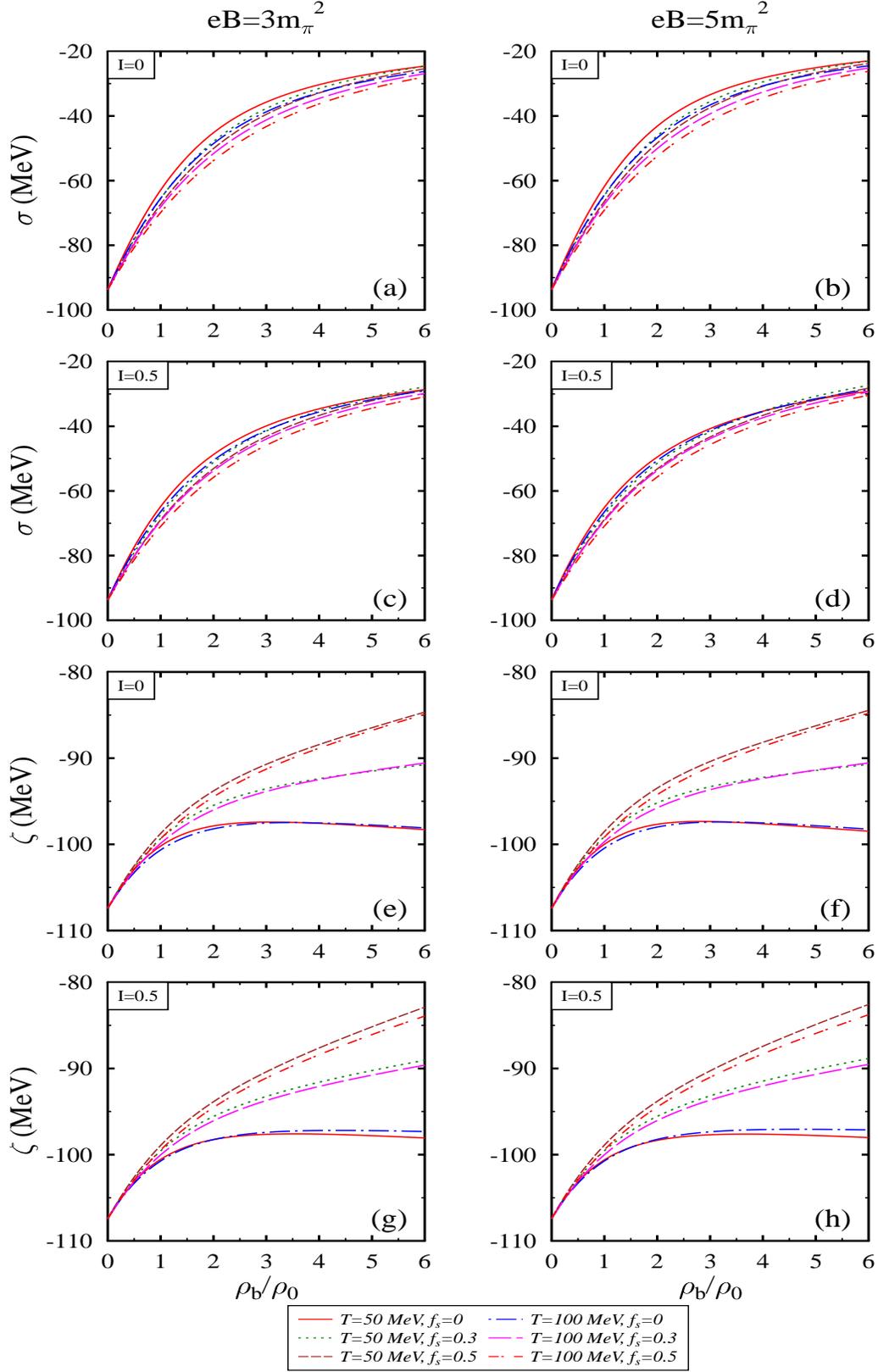}
	\caption{In-medium scalar $\sigma$ and $\zeta$ meson fields}
	\label{sz}
\end{figure}

In the chiral model, the nonlinear coupled equations of motion (\cref{sigma} to \cref{chi}) are solved to calculate the scalar and vector fields at finite density and temperature in the presence of the external magnetic field. The effect of temperature comes into the picture through the thermal distribution function present in the definition of scalar and vector densities, see [\cref{rhovp,rhosp,rhovn,rhosn}]. The AMM for baryons and the sum of Landau energy levels give rise to the effect of magnetic field. In \cref{sz}, we have shown the variation of $\sigma$ and $\zeta$ fields as a function of baryon density ratio ${\rho_b}/{\rho_0}$ at different value of isospin asymmetry ($I$=0 and 0.5) and strangeness fraction ($f_s = 0$, $0.3$ and $0.5$). The results are shown for magnetic field strength $eB=3m_{\pi}^2$ and $5m_{\pi}^2$ at temperature $T$= 50 and 100 MeV.

From subplot (a) to (d) of \cref{sz}, it is observed that the magnitude of $\sigma$ field decreases with an increase in baryonic density for both non-strange ($f_s=0$) and strange medium ($f_s=0.5$). While shifting from $I = 0$ to $0.5$, at a constant magnetic field and finite temperature with zero strangeness fraction, the magnitude of the $\sigma$ field increases. For example, in pure symmetric matter, at $\rho_b={6}{\rho_0}$ and temperature $T = 50 (100)$ MeV, the values of $\sigma$ field are noted as -24.57 (-26.22) and -22.94 (-24.48) MeV for $eB=3m_{\pi}^2$ and $5m_{\pi}^2$ respectively, whereas for isospin asymmetric matter these values changes to -28.67 (-29.00) and -29.17 (-28.65) MeV. We have also noted the variation in magnitude of $\sigma$ field with increased strangeness fraction of the magnetized medium. Considering the above parameters for $f_s=0.5$, the values of $\sigma$ field is found to be -25.39 (-27.93) and -23.76 (-26.19) MeV for $I = 0$ whereas it changes to -28.69 (-30.89) and -28.20 (-30.42) for $I=0.5$. Furthermore, for pure symmetric nuclear matter, increasing the magnetic field from $3m_{\pi}^2$ to $5m_{\pi}^2$, resulted in a modest drop in the magnitude of $\sigma$ field and continues the same nature on taking isospin asymmetric matter into account. Similarly, investigating the field's behavior for a fixed value of magnetic field and asymmetry, it was discovered that its magnitude increases with increasing $f_s$.

For the strange scalar-isoscalar field ($\zeta$), the magnitude of the field decreases with increasing density, at $f_s=0$, and then becomes nearly constant at high baryonic density. However, as we increase the strangeness fraction in the medium, fixing isospin asymmetry and magnetic field, the magnitude of $\zeta$ field always decreases for an increase in baryonic density, as seen from subplot (e) to (h) of \cref{sz}. Contrary to $\sigma$ field, the magnitude of $\zeta$ field decreases with an increase in the strangeness fraction $f_s$ in the medium. This signifies the impact of coupling of $\zeta$ field (which is a strange quark condensate $s\bar{s}$) with hyperons in the medium and this will have significant impact on the $\eta$ meson properties also. The $\zeta$ field shows minor deviation on shifting from isospin symmetric to asymmetric matter.
This impact originates from the non-zero isospin quark content of $\sigma$, whereas $\zeta$ has isospin-independent quark content.

\begin{figure}[ht!] 
	\includegraphics[width=16cm,height=22cm]{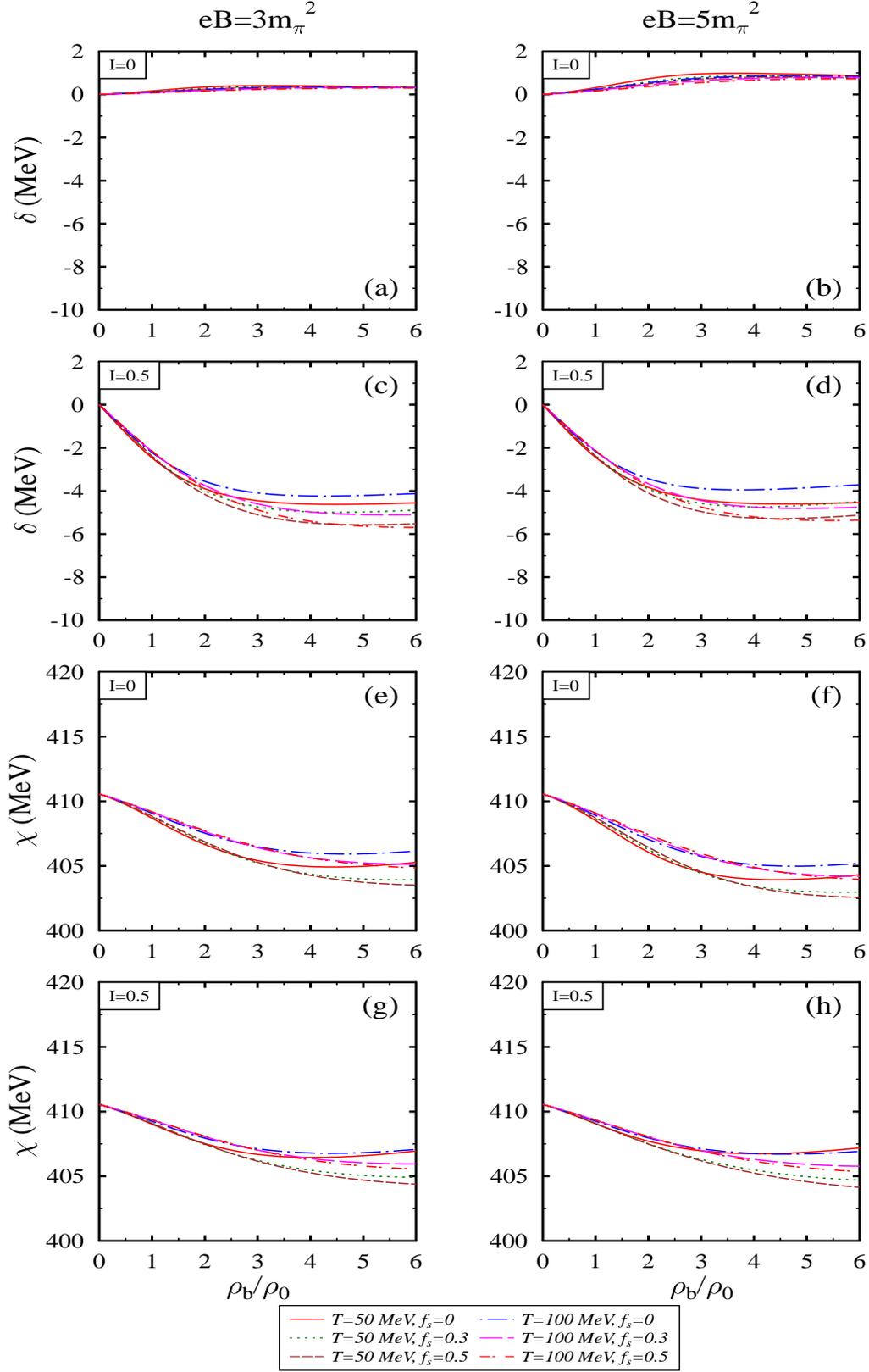}
	\caption{In-medium scalar $\delta$ and $\chi$ meson fields.}
	\label{dc}
\end{figure}

In \cref{dc}, the scalar isovector field ($\delta$) and dilaton field ($\chi$) are plotted as a function of baryonic density along with the contribution of isospin asymmetry and strangeness fraction. At $I=0$ and $eB=0$, the $\delta$ field is zero, but if the magnetic field strength is turned on, the $\delta$ field attains a non-zero value. This is because, in the presence of magnetic field, there is inequality between two different isospin partners in scalar density of baryons, leading to a non-zero value of $\delta$, even though isospin asymmetry parameter $I$ is fixed to zero (\cref{rhovp,rhosn}). In subplots (a) and (b), a small value of $\delta$ $\approx$ 0.3 MeV ($eB = 3m_{\pi}^2$) and $\approx$ 0.8 MeV  ($eB = 5m_{\pi}^2$) is observed for the pure symmetric matter at $\rho_b={6}{\rho_0}$. On the other hand, for the isospin asymmetric case, $i.e.$, $I=0.5$, this behavior of $\delta$ shows an appreciable decrement with density and strangeness. From the subplots (c) and (d) of \cref{dc}, we observe the magnitude of $\delta$ field for $eB=3m_{\pi}^2 (5m_{\pi}^2)$, at $f_s=0$, $T=50$ MeV, and $\rho_b={6}\rho_0$ as $4.55 (4.53)$ MeV, which changes to $5.52 (5.12)$ for $f_s=0.5$. In addition, we have plotted the glueball field $\chi$, which incorporates the scale invariance property of QCD \cite{Papazoglou1999}. It is seen that the Lagrangian terms ${\cal L} _{0}$ (\cref{lagscal}) and ${\cal L} _{SB}$ (\cref{lsb}) are responsible for coupling of $\chi$ with $\sigma$, $\zeta$ and $\delta$. From \cref{dc}, we can observe a decrease in $\chi$ with increasing baryonic density. Also, a significant magnetic field effect is observed in the high-density regime. 
The coupling of $\chi$ with other scalar fields causes a decrease in its magnitude with increasing $f_s$. For example, considering the case of isospin asymmetric matter, at $T=50$ MeV, $\rho_b={6}{\rho_0}$ and $eB=3m_{\pi}^2$ ($5m_{\pi}^2$), the $\chi$ posses the value 406.95 (407.19) MeV for $f_s=0$ which decreases to 404.39 (404.14) MeV for $f_s=0.5$. Compared with nuclear matter, this field shows the least variation among other scalar fields, leading to the assumption of frozen glueball limit in literature \cite{Mishra2004,Reddy18,Zschiesche}.  


\begin{figure}[ht!]
	\includegraphics[width=16cm,height=22cm]{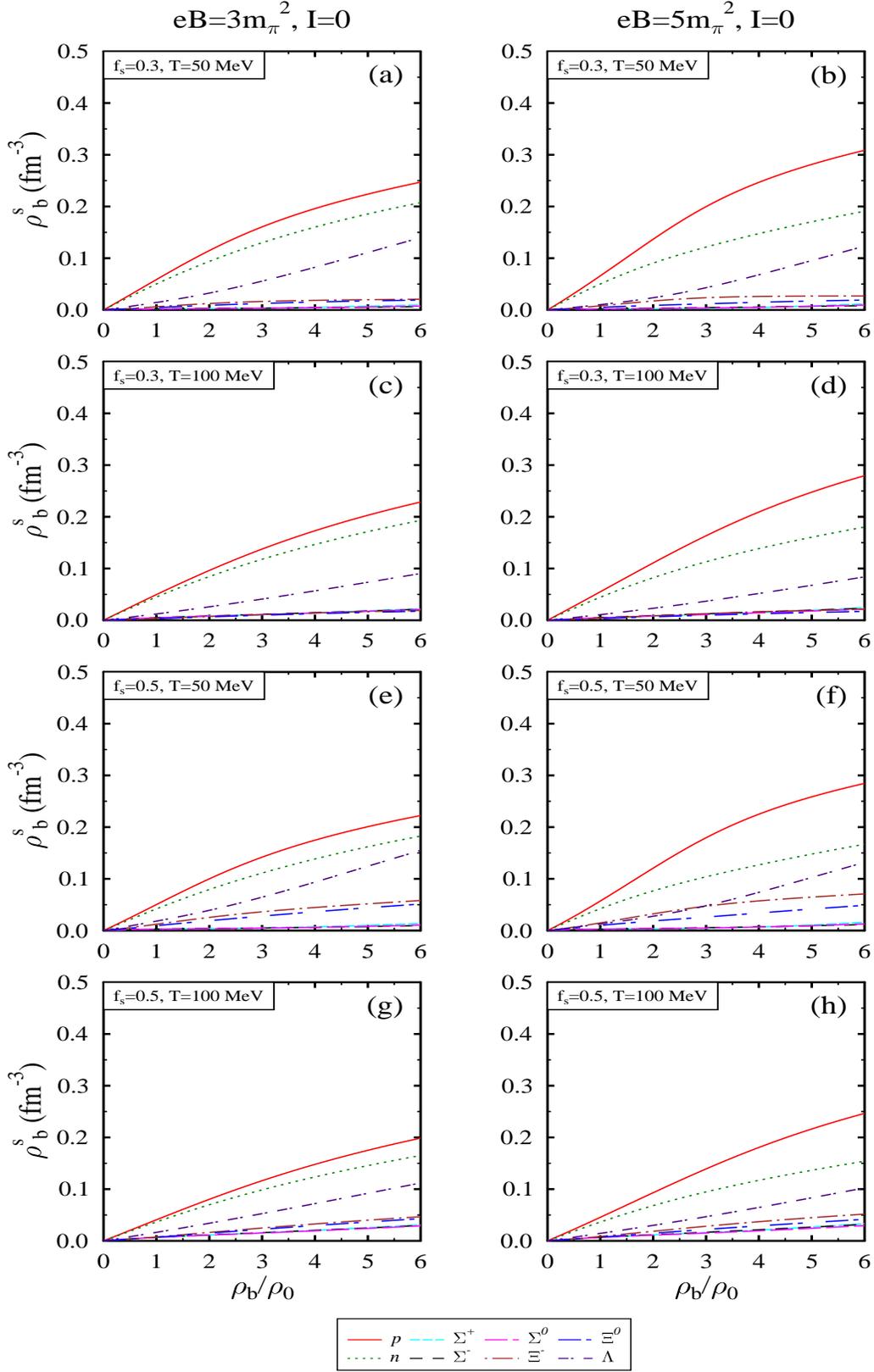}
	\caption{In-medium scalar density of baryons for pure symmetric matter $I=0$.}
	\label{sden0}
\end{figure}
\begin{figure}[ht!]
	\includegraphics[width=16cm,height=22cm]{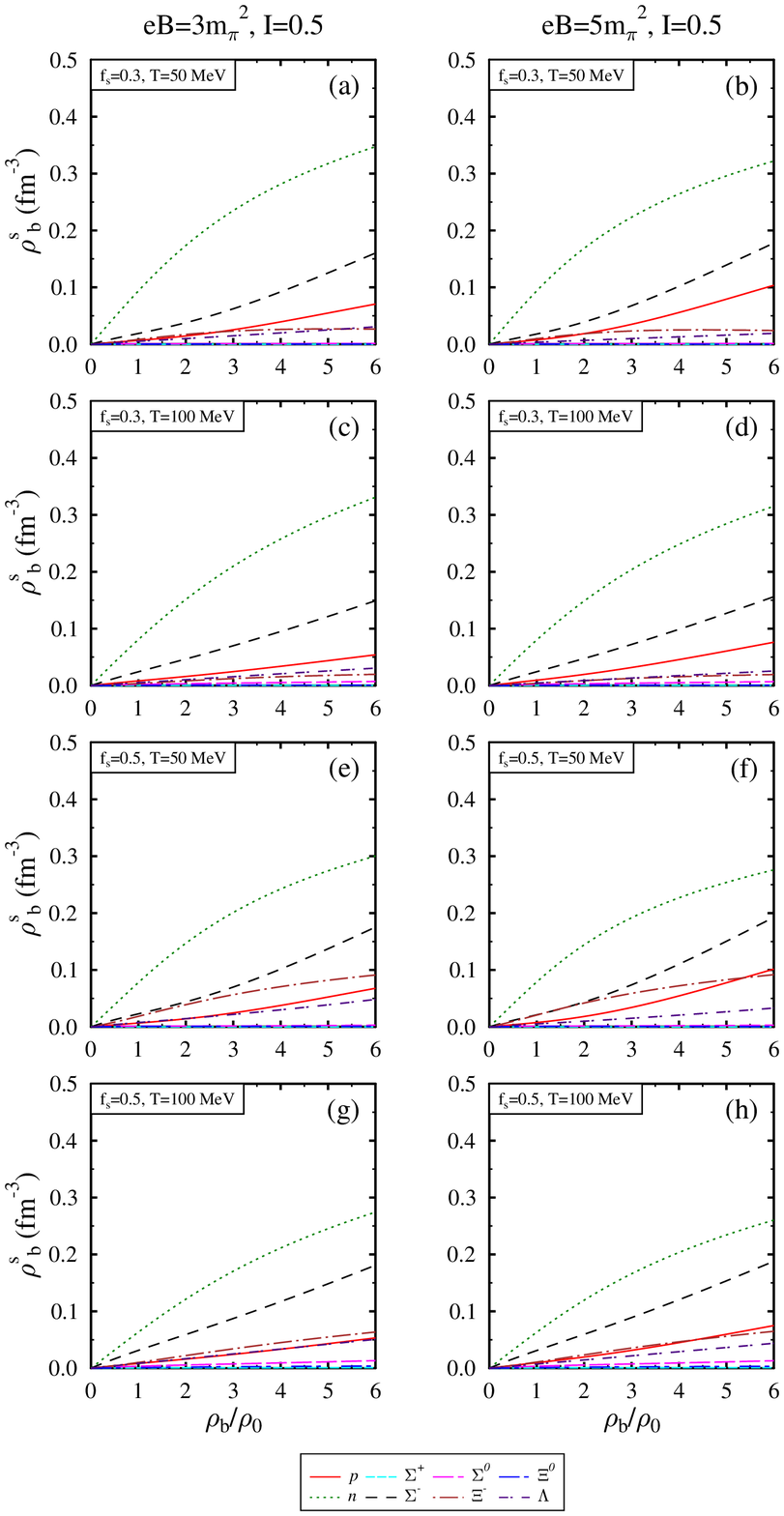}
	\caption{In-medium scalar density of baryons for isospin asymmetric matter $I=0.5$.}
	\label{sden5}
\end{figure}

In \cref{sden0,sden5}, the scalar density of baryons ($p, n, \Sigma^+, \Sigma^0, \Sigma^-, \Xi^0$, $\Xi^-, \Lambda$) is plotted as a function of baryonic density for finite values of temperature, magnetic field, and isospin asymmetry ($I=0$ and $0.5$). In addition, the effect of strangeness is also taken into account. Here, a significant temperature impact is observed in the strange matter for higher densities. We noted a decrease in scalar densities as a function of temperature due to the coupled equations and the Fermi distribution integral \cite{Rajesh2020eta}. This impact is more appreciable considering the higher value of the magnetic field. In \cref{sden0,sden5}, an increase in scalar density is observed, which becomes more pronounced on changing $eB$ from $3m_{\pi}^2$ to $5m_{\pi}^2$. In the case of nuclear matter, $i.e.$, $f_s=0$, all other baryon densities vanish except $\rho_p$ and $\rho_n$. For zero isospin asymmetry and magnetic field, there is no symmetry breaking in $\rho_p$ and $\rho_n$ therefore they show the same values, but when the magnetic field is turned on, a splitting is observed in $\rho_p$ and $\rho_n$ despite being $I=0$. This happens due to the additional landau effects in protons. Moving from $I=0$ to $0.5$, the plot shows an increase in neutron scalar density, whereas the proton scalar density decreases. This happens due to the non-zero contribution of $\delta$ and $\rho$ fields in isospin asymmetric matter, resulting to the change in effective mass and chemical potential and thus, the change in scalar density. 


\subsection{$\eta \cal{B}$ medium-modified mass and Optical Potential} 
\label{rs III B}
\begin{figure}[ht!]
	\includegraphics[width=16cm,height=22cm]{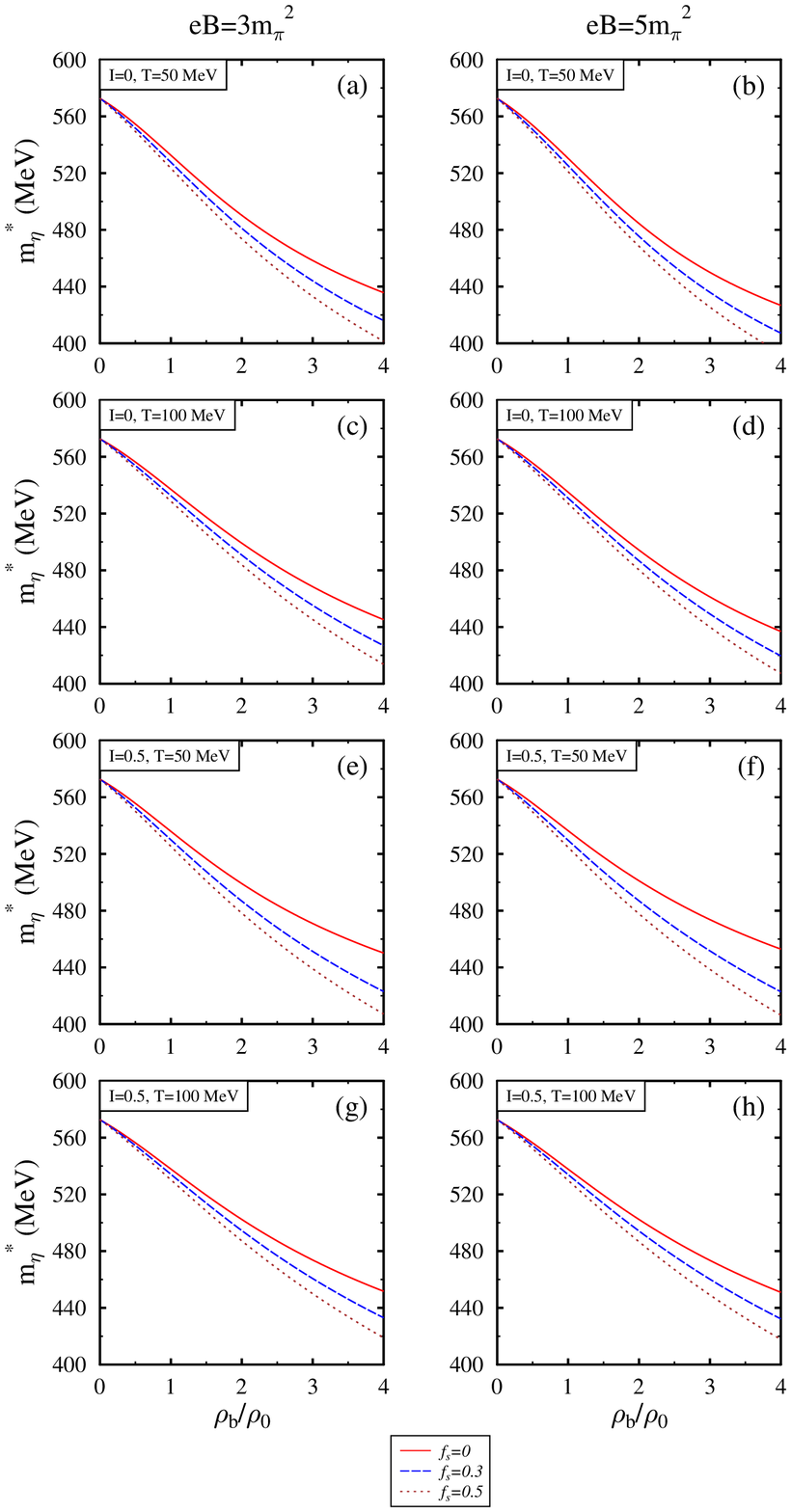}
	\caption{ In-medium $\eta$ meson mass for $a^{\eta N} = 0.91$ fm.}
	\label{masseta0.91}
\end{figure}

\begin{figure}[ht!]
	\includegraphics[width=16cm,height=22cm]{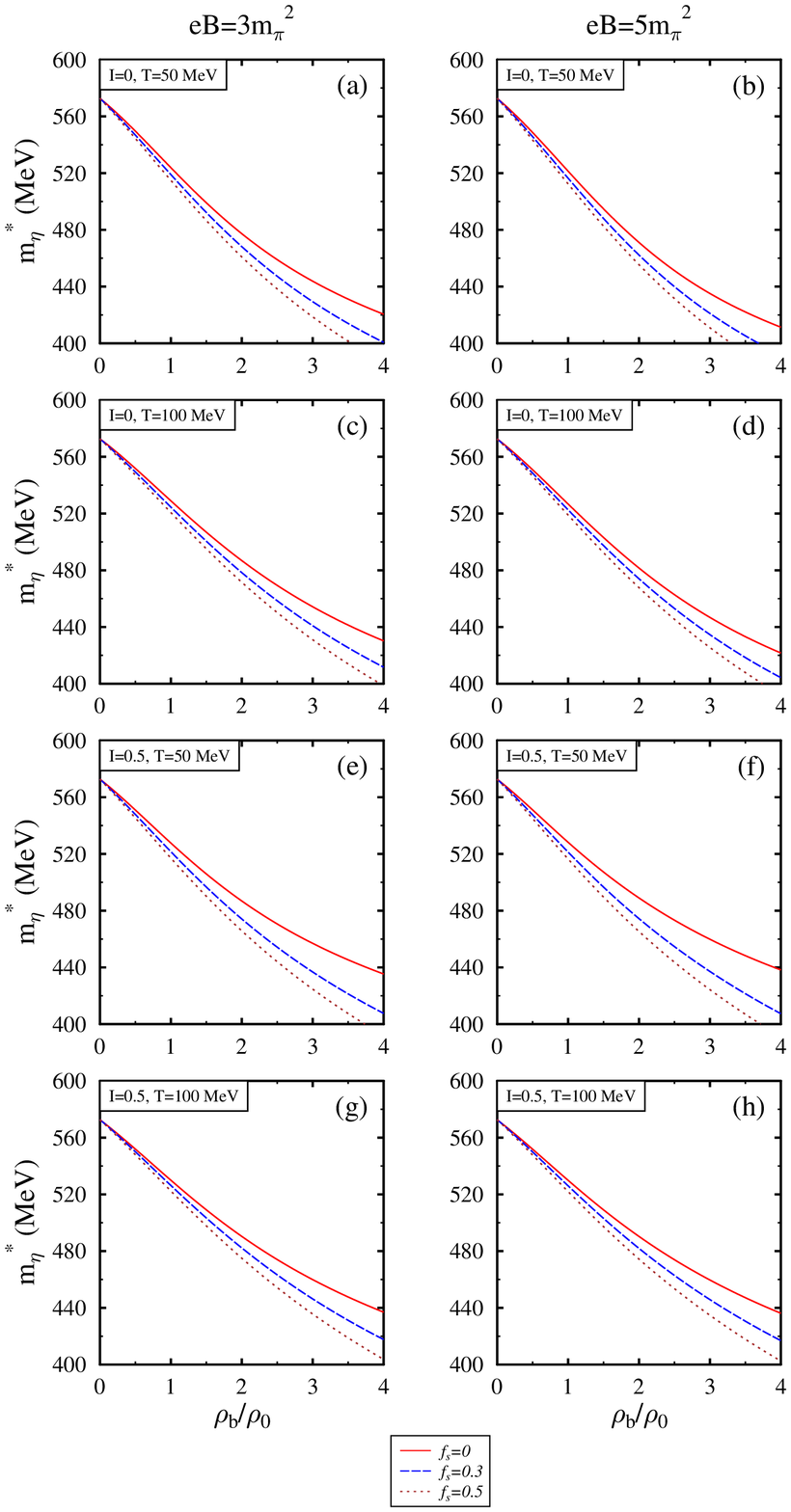}
	\caption{ In-medium $\eta$ meson mass for $a^{\eta N} = 1.02$ fm.}
	\label{masseta1.02}
\end{figure}

\begin{figure}[ht!]
	\includegraphics[width=16cm,height=22cm]{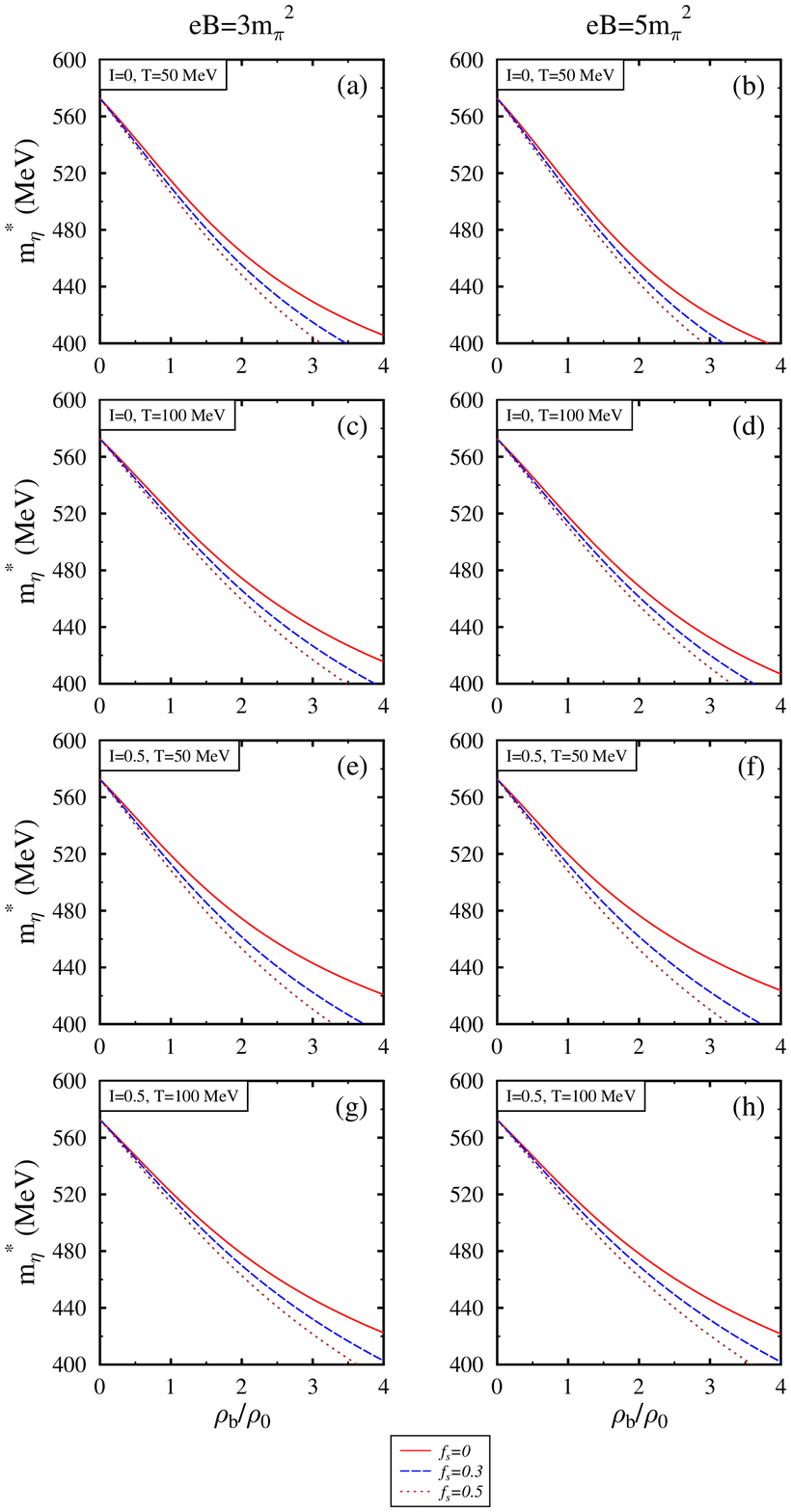}
	\caption{ In-medium $\eta$ meson mass for $a^{\eta N} = 1.14$ fm.}
	\label{masseta1.14}
\end{figure}

In \cref{masseta0.91,masseta1.02,masseta1.14}, the medium modified mass of $\eta$ meson is illustrated as a function of baryonic density at different values of temperature, magnetic field, and for various scattering lengths ($a^{\eta N}=0.91$, $1.02$ and $1.14$ fm) \cite{Zhong2006}. The in-medium mass is obtained through the solution of the dispersion relation [see \cref{drk}]. In the present calculation, we have used the parameters $d_1$=$2.56/m_K$ and $d_2$=$0.73/m_K$ obtained from kaon-nucleon scattering lengths \cite{Mazumdar2009, Mishra2019a}. Considering the effects of temperature, isospin, and strangeness, we observed a decrease in the in-medium mass ($m_\eta^*$) as a function of baryonic density. A linear decrement is observed in the low-density regime. In contrast, a non-linearity is noticed in the high-density regime. This behavior is opposite to the baryon scalar density of \cref{sden0,sden5}. This is due to the dependency of $\eta$ meson self-energy ($\Pi^*$) on baryon scalar densities [\cref{sen}].

With an increase in the scattering length, $a^{\eta N}$ from $0.91$ to $1.14$ fm, the effective mass decreases due to the direct dependence of $d^\prime$ term on $a^{\eta N}$ [\cref{sl}]. On increasing the scattering length, the value of $d^\prime$ also increases. As the self-energy has an attractive contribution concerning the $d^\prime$ term, a decrease in effective mass is observed. For example, in the case of pure symmetric nuclear matter at $eB=3m_{\pi}^2$ and $T=50$ MeV, the effective mass changes from 532.59 (435.71) to 514.78 (405.60) MeV, for $\rho_b = \rho_0 (4\rho_0)$ (\cref{tabm3eBfs0}). On analyzing the three terms of \cref{sen}, we can observe that the first range term contributed repulsively, whereas the other two terms contributed attractively. The $d^\prime$ term was dominant and significantly contributed to the final effective mass. From \cref{masseta0.91,masseta1.02,masseta1.14}, we have observed a rapid fall in effective mass at the high-density regime. This observation signifies the notable role of $a^{\eta N}$ in $\eta \mathcal{B}$ interactions.
 \begin{table}[ht!]
\begin{tabular}{|c|c|c|c|c|c|c|c|c|c|}
\hline
&& \multicolumn{4}{c|}{I=0, $f_s$=0}    & \multicolumn{4}{c|}{I=0.5, $f_s$=0}   \\
\cline{3-10}
&$a^{\eta N} (\text{fm})$ & \multicolumn{2}{c|}{T=50} & \multicolumn{2}{c|}{T=100 }& \multicolumn{2}{c|}{T=50}& \multicolumn{2}{c|}{T=100 }\\
\cline{3-10}
&  &$\rho_0$&$4\rho_0$ &$\rho_0$  &$4\rho_0$ & $\rho_0$ &$4\rho_0$&$\rho_0$&$4\rho_0$ \\ \hline 
& 0.91 & 532.59 & 435.71 & 537.04 & 445.19 & 535.93 & 450.15 & 537.86 & 451.67 \\ \cline{2-10}
$ m^*_\eta$& 1.02 & 523.84 & 420.50 & 528.96 & 430.22 & 527.69 & 435.37 & 529.98 & 436.95 \\ \cline{2-10}
&1.14 & 514.78 & 405.60 & 520.56 & 415.50 & 519.11 & 420.80 & 521.77 & 422.42\\ \cline{1-10}

\end{tabular}
\caption{The tabulated values of $\eta$ meson in-medium mass for $eB=3m_{\pi}^2$ and $f_s=0$ (in MeV).}
\label{tabm3eBfs0}
\end{table}
 \begin{table}[ht!]
\begin{tabular}{|c|c|c|c|c|c|c|c|c|c|}
\hline
&& \multicolumn{4}{c|}{I=0, $f_s$=0.3}    & \multicolumn{4}{c|}{I=0.5, $f_s$=0.3}   \\
\cline{3-10}
&$a^{\eta N} (\text{fm})$ & \multicolumn{2}{c|}{T=50} & \multicolumn{2}{c|}{T=100 }& \multicolumn{2}{c|}{T=50}& \multicolumn{2}{c|}{T=100 }\\
\cline{3-10}
&  &$\rho_0$&$4\rho_0$ &$\rho_0$  &$4\rho_0$ & $\rho_0$ &$4\rho_0$&$\rho_0$&$4\rho_0$ \\ \hline 
& 0.91 & 527.63 & 416.04 & 532.58 & 426.90 & 529.96 & 423.00 & 534.19 & 432.99 \\ \cline{2-10}
$ m^*_\eta$& 1.02 & 518.99 & 400.69 & 524.53 & 411.67 & 521.62 & 407.61 & 526.28 & 417.66 \\ \cline{2-10}
&1.14 & 510.04 & 385.73 & 516.15 & 396.79 & 512.95 & 392.59 & 518.04 & 402.67\\ \cline{1-10}

\end{tabular}
\caption{The tabulated values of $\eta$ meson in-medium mass for $eB=3m_{\pi}^2$ and $f_s=0.3$ (in MeV).}
\label{tabm3eBfs3}
\end{table}

We observed a small increase in the effective mass of $\eta$ mesons with increase of temperature. Additionally, the $\eta$ meson experiences a decrease in effective mass values on increasing strangeness at the different magnetic field. This decrement of mass on changing $f_s=0$ to $0.3$ at constant magnetic field (say $3m_{\pi}^2$) results from interaction of hyperons with strange $\zeta$ field. As was discussed earlier,  $\zeta$ field contribute to considerable attractive interactions in the medium as a function of $f_s$ and this causes a decrease in the effective mass of $\eta$ mesons, as can be seen from \cref{tabm3eBfs0,tabm3eBfs3}. Also, on switching to a higher magnetic field ($5m_{\pi}^2$), for given isospin asymmetry of the medium, a total decrement is observed in effective mass, (compare \cref{tabm3eBfs0,tabm5eBfs0} and \cref{tabm3eBfs3,tabm5eBfs3}). 

 \begin{table}
\begin{tabular}{|c|c|c|c|c|c|c|c|c|c|}
\hline
& & \multicolumn{4}{c|}{I=0, $f_s$=0}    & \multicolumn{4}{c|}{I=0.5, $f_s$=0}   \\
\cline{3-10}
&$a^{\eta N} (\text{fm})$ & \multicolumn{2}{c|}{T=50} & \multicolumn{2}{c|}{T=100 }& \multicolumn{2}{c|}{T=50}& \multicolumn{2}{c|}{T=100 }\\
\cline{3-10}
&  &$\rho_0$&$4\rho_0$ &$\rho_0$  &$4\rho_0$ & $\rho_0$ &$4\rho_0$&$\rho_0$&$4\rho_0$ \\ \hline 
& 0.91 & 530.44 & 426.74 & 535.09 & 436.86 & 536.41 & 452.91 & 537.80 & 450.93 \\ \cline{2-10}
$ m^*_\eta$& 1.02 & 521.39 & 411.35 & 526.72 & 421.68 & 528.24 & 438.23 & 529.92 & 436.18 \\ \cline{2-10}
& 1.14 & 512.03 & 396.34 & 518.03 & 406.80 & 519.74 & 423.74 & 521.70 & 421.62\\ \cline{1-10}

\end{tabular}
\caption{The tabulated values of $\eta$ meson in-medium mass for $eB=5m_{\pi}^2$ and $f_s=0$ (in MeV).}
\label{tabm5eBfs0}
\end{table}

 \begin{table}
\begin{tabular}{|c|c|c|c|c|c|c|c|c|c|}
\hline
& & \multicolumn{4}{c|}{I=0, $f_s$=0.3}    & \multicolumn{4}{c|}{I=0.5, $f_s$=0.3}   \\
\cline{3-10}
&$a^{\eta N} (\text{fm})$ & \multicolumn{2}{c|}{T=50} & \multicolumn{2}{c|}{T=100 }& \multicolumn{2}{c|}{T=50}& \multicolumn{2}{c|}{T=100 }\\
\cline{3-10}
&  &$\rho_0$&$4\rho_0$ &$\rho_0$  &$4\rho_0$ & $\rho_0$ &$4\rho_0$&$\rho_0$&$4\rho_0$ \\ \hline 
& 0.91 & 525.30 & 407.22 & 530.94 & 419.64 & 529.66 & 422.82 & 533.98 & 432.24 \\ \cline{2-10}
$ m^*_\eta$& 1.02 & 516.45 & 391.85 & 522.63 & 404.29 & 521.34 & 407.39 & 526.06 & 416.88 \\ \cline{2-10}
& 1.14 & 507.28 & 376.91 & 514.02 & 389.33 & 512.69 & 392.34 & 517.80 & 401.86\\ \cline{1-10}

\end{tabular}
\caption{The tabulated values of $\eta$ meson in-medium mass for $eB=5m_{\pi}^2$ and $f_s=0.3$ (in MeV).}
\label{tabm5eBfs3}
\end{table}


 \begin{table}
\begin{tabular}{|c|c|c|c|c|c|c|c|c|c|}
\hline
& & \multicolumn{4}{c|}{I=0, $f_s$=0}    & \multicolumn{4}{c|}{I=0.5, $f_s$=0}   \\
\cline{3-10}
&$a^{\eta N} (\text{fm})$ & \multicolumn{2}{c|}{T=50} & \multicolumn{2}{c|}{T=100 }& \multicolumn{2}{c|}{T=50}& \multicolumn{2}{c|}{T=100 }\\
\cline{3-10}
&  &$k_{00}$&$k_{10}$ &$k_{00}$  &$k_{10}$ & $k_{00}$ &$k_{10}$&$k_{00}$&$k_{10}$ \\ \hline 
& 0.91 & -41.78 & -20.25 & -37.34 & -18.15 & -38.45 & -18.67 & -36.52 & -17.76 \\ \cline{2-10}
$ U^*_\eta$& 1.02 & -50.53 & -24.34 & -45.41 & -21.95 & -46.69 & -22.55 & -44.39 & -21.47 \\ \cline{2-10}
& 1.14 & -59.59 & -28.52 & -53.81 & -25.86 & -55.26 & -26.52 & -52.60 & -25.29\\ \cline{1-10}

\end{tabular}
\caption{The values of optical potential of $\eta$ meson for $eB=3m_{\pi}^2$ (in MeV) and $f_s=0$. Here $k_{00}$ and $k_{10}$ represent the value of momentum at $|\textbf{k}|=0$ and $|\textbf{k}|=1000$ MeV respectively.}
\label{tabopt3eBfs0}
\end{table}

 \begin{table}
\begin{tabular}{|c|c|c|c|c|c|c|c|c|c|}
\hline
& & \multicolumn{4}{c|}{I=0, $f_s$=0.3}    & \multicolumn{4}{c|}{I=0.5, $f_s$=0.3}   \\
\cline{3-10}
&$a^{\eta N} (\text{fm})$ & \multicolumn{2}{c|}{T=50} & \multicolumn{2}{c|}{T=100 }& \multicolumn{2}{c|}{T=50}& \multicolumn{2}{c|}{T=100 }\\
\cline{3-10}
&  &$k_{00}$&$k_{10}$ &$k_{00}$  &$k_{10}$ & $k_{00}$ &$k_{10}$&$k_{00}$&$k_{10}$ \\ \hline 
& 0.91 & -46.75 & -22.57 & -41.79 & -20.25 & -44.41 & -21.48 & -40.18 & -19.49 \\ \cline{2-10}
$ U^*_\eta$& 1.02 & -55.38 & -26.58 & -49.84 & -24.01 & -52.76 & -25.37 & -48.09 & -23.20 \\ \cline{2-10}
& 1.14 & -64.34 & -30.68 & -58.23 & -27.89 & -61.42 & -29.35 & -56.33 & -27.02\\ \cline{1-10}

\end{tabular}
\caption{The values of optical potential of $\eta$ meson for $eB=3m_{\pi}^2$ (in MeV) and $f_s=0.3$.}
\label{tabopt3eBfs3}
\end{table}

 \begin{table}
\begin{tabular}{|c|c|c|c|c|c|c|c|c|c|}
\hline
& & \multicolumn{4}{c|}{I=0, $f_s$=0}    & \multicolumn{4}{c|}{I=0.5, $f_s$=0}   \\
\cline{3-10}
&$a^{\eta N} (\text{fm})$ & \multicolumn{2}{c|}{T=50} & \multicolumn{2}{c|}{T=100 }& \multicolumn{2}{c|}{T=50}& \multicolumn{2}{c|}{T=100 }\\
\cline{3-10}
&  &$k_{00}$&$k_{10}$ &$k_{00}$  &$k_{10}$ & $k_{00}$ &$k_{10}$&$k_{00}$&$k_{10}$ \\ \hline  
& 0.91 & -43.93 & -21.26 & -39.28 & -19.07 & -37.96 & -18.44 & -36.57 & -17.79 \\ \cline{2-10}
$ U^*_\eta$& 1.02 & -52.98 & -25.47 & -47.65 & -22.99 & -46.13 & -22.28 & -44.45 & -21.50 \\ \cline{2-10}
& 1.14 & -62.35 & -29.77 & -56.35 & -27.02 & -54.63 & -26.23 & -52.67 & -25.33\\ \cline{1-10}
\end{tabular}
\caption{The values of optical potential of $\eta$ meson for $eB=5m_{\pi}^2$ (in MeV) and $f_s=0$.}
\label{tabopt5eBfs0}
\end{table}
 \begin{table}
\begin{tabular}{|c|c|c|c|c|c|c|c|c|c|}
\hline
& & \multicolumn{4}{c|}{I=0, $f_s$=0.3}    & \multicolumn{4}{c|}{I=0.5, $f_s$=0.3}   \\
\cline{3-10}
&$a^{\eta N} (\text{fm})$ & \multicolumn{2}{c|}{T=50} & \multicolumn{2}{c|}{T=100 }& \multicolumn{2}{c|}{T=50}& \multicolumn{2}{c|}{T=100 }\\
\cline{3-10}
&  &$k_{00}$&$k_{10}$ &$k_{00}$  &$k_{10}$ & $k_{00}$ &$k_{10}$&$k_{00}$&$k_{10}$ \\ \hline 
& 0.91 & -49.06 & -23.65 & -43.47 & -21.04 & -44.72 & -21.62 & -40.39 & -19.59 \\ \cline{2-10}
$ U^*_\eta$& 1.02 & -57.93 & -27.75 & -51.75 & -24.89 & -53.03 & -25.49 & -48.32 & -23.30 \\ \cline{2-10}
& 1.14 & -67.09 & -31.93 & -60.35 & -28.86 & -61.68 & -29.47 & -56.57 & -27.13\\ \cline{1-10}

\end{tabular}
\caption{The values of optical potential of $\eta$ meson for $eB=5m_{\pi}^2$ (in MeV) and $f_s=0.3$.}
\label{tabopt5eBfs3}
\end{table}

\begin{figure}[ht!]
	\includegraphics[width=16cm,height=22cm]{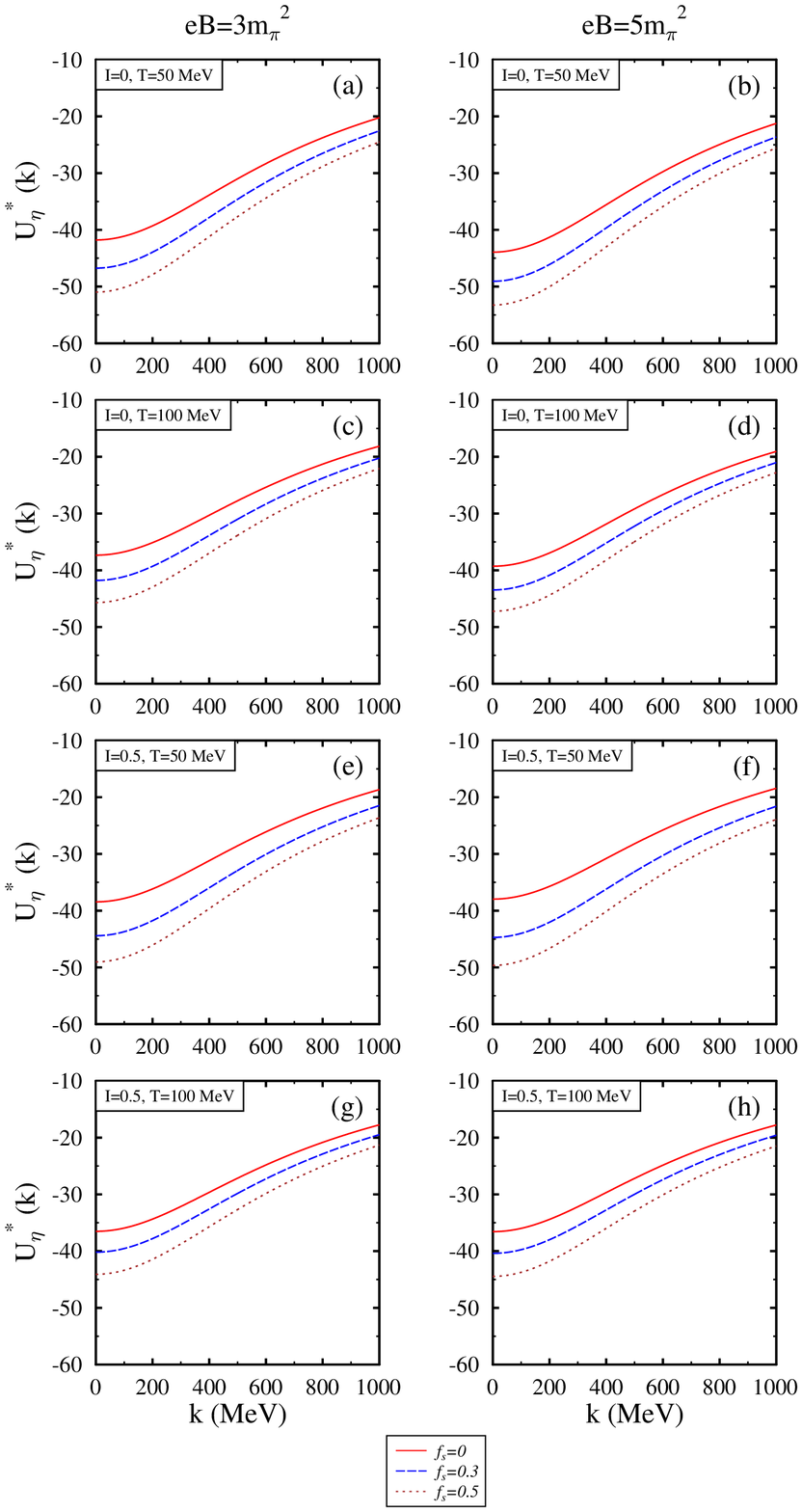}
	\caption{ In-medium $\eta$ meson optical potential for $a^{\eta N} = 0.91$ fm.}
	\label{opteta0.91}
\end{figure}

\begin{figure}[ht!]
	\includegraphics[width=16cm,height=22cm]{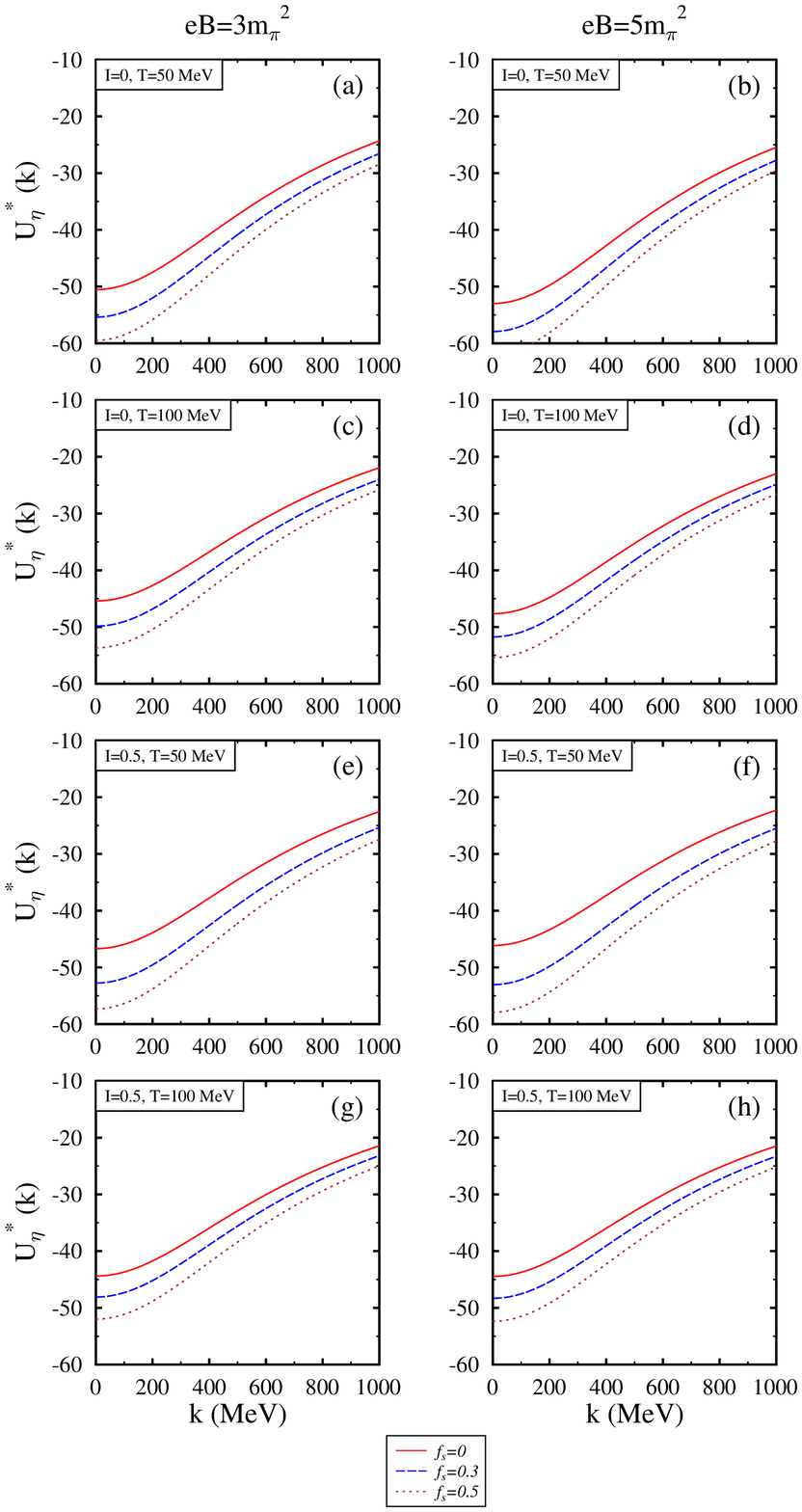}
	\caption{ In-medium $\eta$ meson optical potential for $a^{\eta N} = 1.02$ fm.}
	\label{opteta1.02}
\end{figure}

\begin{figure}[ht!]
	\includegraphics[width=16cm,height=22cm]{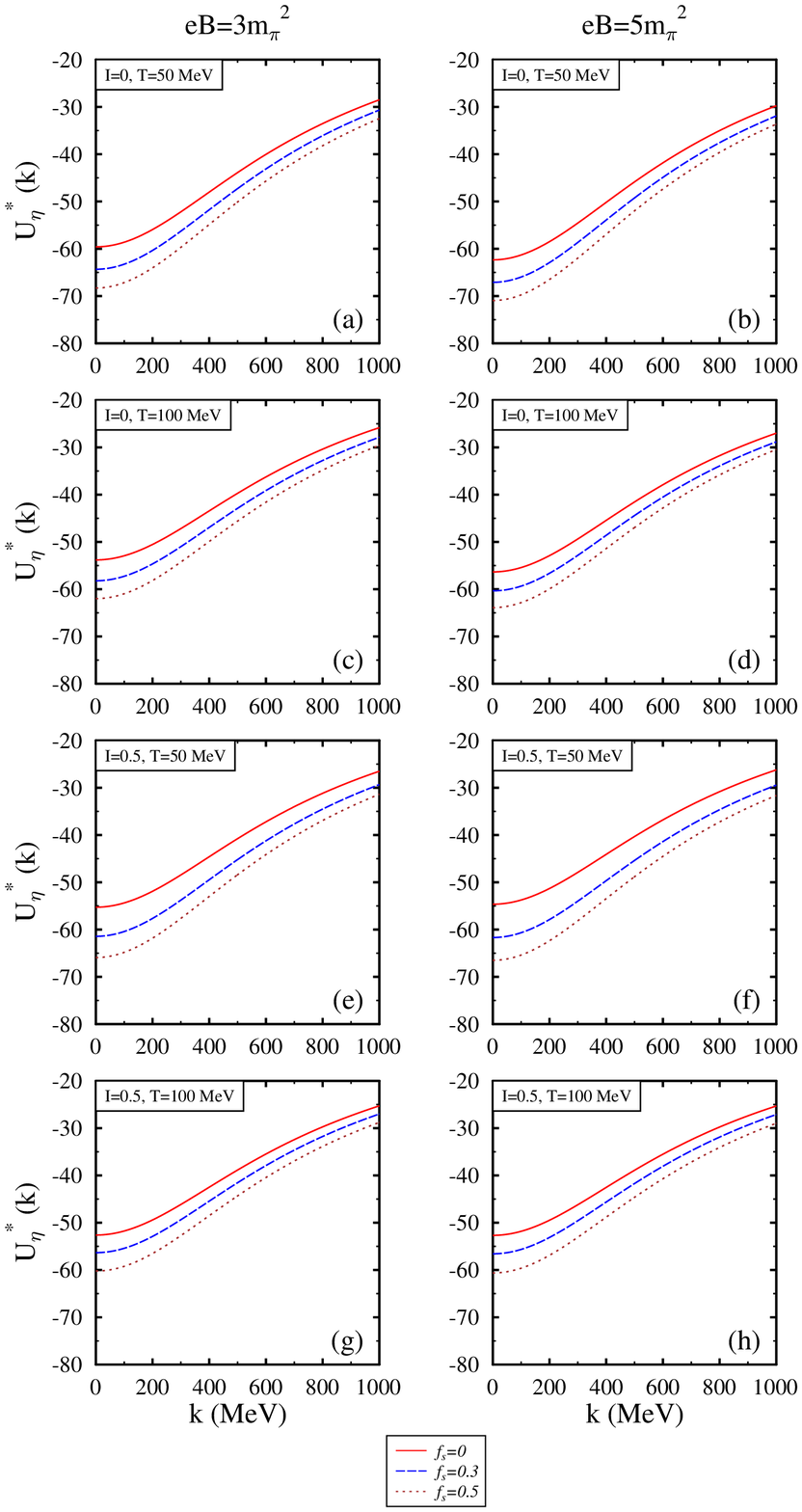}
	\caption{ In-medium $\eta$ meson optical potential for $a^{\eta N} = 1.14$ fm.}
	\label{opteta1.14}
\end{figure}

The possibility of $\eta$ meson bound state arises with decreasing in-medium mass \cite{Zhong2006,Cieply14}. Therefore, the optical potential of $\eta$, $U^*_{\eta}$ is studied in chiral SU(3) hadronic mean-field model, which may be useful for understanding the bound states. We have plotted $U^*_{\eta}$ with respect to momentum $|\textbf{k}|$ involving different parameters in \cref{opteta0.91,opteta1.02,opteta1.14}. It is observed that the behavior of $U^*_{\eta}$ is getting less negative with increasing momentum. Moreover, \cref{opk} signifies the momentum dominance for higher values of $|\textbf{k}|$ resulting in the small contribution of effective mass. Furthermore, $U^*_{\eta}$ is strongly influenced by scattering length $a^{\eta N}$. When we modify the value of $a^{\eta N}$ from 0.91 $\sim$ 1.14 fm, for pure symmetric nuclear matter, the optical potential changes from -41.78 (-20.25) to -59.59 (-28.52) for $k = 0 (1000)$ MeV at magnetic field $eB=3m_{\pi}^2$, as provided in \cref{tabopt3eBfs0}. On changing the magnetic field for given strangeness and isospin asymmetry, a small increase in the magnitude of optical potential is observed, which is tabulated in \cref{tabopt3eBfs3,tabopt5eBfs3}. Further, if we change the strangeness fraction for a given value of isospin asymmetry, scattering length, and magnetic field, a notable increase in the magnitude of optical potential is observed. For example, keeping $a^{\eta N} =0.91$ fm, $I=0$, and $eB=5m_{\pi}^2$, at $|\textbf{k}| = 0 (1000)$ MeV, the magnitude of $U^*_{\eta}$ is observed to be 43.93 (21.26) MeV for $f_s=0$, which increases to 49.06 (23.65) MeV at $f_s=0.3$ (\cref{tabopt5eBfs0,tabopt5eBfs3}). Also, a change in temperature from $T=50$ to $100$ MeV shows a small effect in the values of $U^*_{\eta}$.

To summarize, from the present calculations we observe that the inclusion of hyperons along with nucleons have significant impact on the in-medium masses and optical potentials of $\eta$ mesons. The optical potential of mesons can be utilized for investigating the bound states known as $\eta$-mesic nuclei. Further, these in-medium calculations can be used as input in various transport approaches to understand the experimental observables expected to produce in future experimental facilities where compressed baryonic matter at significant high densities may be created (for example, CBM experiment of FAIR \cite{CBM22}, J-PARC \cite{JPARC2022}, and NICA \cite{NICA2022}).

\end{document}